\documentclass[12pt]{article}
\pdfoutput=1  
\newcommand{\TITLE}{Dependence of elastic hadron collisions on impact parameter}
\newcommand{\KEYWORDS}{elastic scattering of hadrons, impact parameter, central or peripheral scattering, proton-proton collisions, Coulomb-hadronic interaction}

\RequirePackage[T1]{fontenc}

\usepackage{amssymb,amsmath,latexsym}
\usepackage[a4paper, twoside,
            top=2.cm,bottom=2.cm,
            left=1.6cm,right=1.6cm,
            bindingoffset=0.6cm
            ]{geometry}

\RequirePackage{graphicx}
\RequirePackage{mathptmx}      
\RequirePackage{flushend}
\RequirePackage[numbers,sort&compress]{natbib}

\usepackage{sidecap}
\usepackage{ifthen}
\usepackage{keycommand}
\usepackage[toc,page]{appendix}
\usepackage[labelfont={small,bf},textfont=small]{caption}
\usepackage[labelfont={small,bf},textfont=small]{subcaption}

\usepackage{placeins} 

\makeatletter \let\cl@chapter\relax \makeatother 

\usepackage{hyperref}
\hypersetup{
    pdftex,
    bookmarks=true,         
    bookmarksnumbered=true,
    bookmarksopen=true,
    bookmarksopenlevel=0,
    bookmarksnumbered=true,
    pdftitle={\TITLE},    
    pdfauthor={J. Prochazka, M. V. Lokajicek, V. Kundrat},      
    pdfsubject={HEP-theory}, 
    pdfkeywords={\KEYWORDS}, 
}

\usepackage{cleveref}  
\newcommand{\refp}[1]{(\ref{#1})}

\crefname{equation}{Eq.}{Eqs.}
\crefname{section}{Sec.}{Secs.}
\crefname{chapter}{Chapter}{Chapters}
\crefname{table}{Table}{Tables}
\crefname{figure}{Fig.}{Figs.}
\crefname{appsec}{Appendix}{Appendices}
\crefname{appchap}{Appendix}{Appendices}

\renewcommand\Re{\operatorname{Re}}
\renewcommand\Im{\operatorname{Im}}
\DeclareMathOperator{\e}{e}
\makeatletter
\def\blfootnote{\xdef\@thefnmark{}\@footnotetext}
\makeatother

\newcommand{\abs}[1]{\ensuremath{\left| {#1} \right|}}
\newcommand{\ampl}[2][]{\ensuremath{F_{#1}^{\text{#2}}(s,t)}}
\newcommand{\modulus}[2][]{\ensuremath{\abs{\ampl[#1]{#2}}}}
\newcommand{\phase}[1][]{\ensuremath{\zeta^\text{N}_{#1}(s,t)}}

\newcommand{\PROF}[1]{\ensuremath{D^{\text{#1}}}}

\newcommand{\dcss}[3][]{
\ifthenelse{\equal{#1}{}}
{\ensuremath{      \frac{\text{d}\sigma^{#2}{#3}}{\text{d}t}}}
{\ensuremath{\left.\frac{\text{d}\sigma^{#2}{#3}}{\text{d}t}\right|_{#1}}}
}

\newcommand{\dcs}[2][]{\dcss[#1]{#2}{}}

\newcommand{\weight}[0]{w}
\newkeycommand{\meanb}[n=1,etype=,j=,weight=]{\ensuremath{\langle \ifthenelse{\equal{\commandkey{n}}{1}}{b}{b^{\commandkey{n}}} \rangle^{\text{\commandkey{etype}}\ifthenelse{\equal{\commandkey{weight}}{}}{}{,{\text{\weight}=\commandkey{weight}}}}_{\commandkey{j}} }}
\newkeycommand{\bmax}[j=]{ \ensuremath{b^{\text{max}}\ifcommandkey{j}{_{\commandkey{j}}}{} }}

\newkeycommand{\PROB}[etype=,j=]{ \ensuremath{\ifcommandkey{etype}{P^{\text{\commandkey{etype}}}}{P}_{\ifcommandkey{j}{\commandkey{j}}{}} }}

\newkeycommand{\CS}[etype=,j=]{ \ensuremath{\ifcommandkey{etype}{\sigma^{\text{\commandkey{etype}}}}{\sigma}_{\ifcommandkey{j}{\commandkey{j}}{}} }}

\newkeycommand{\dbt}[j=]{ \ensuremath{ d_{b\ifthenelse{ \equal{\commandkey{j}}{} }{}{,\commandkey{j}}}(t) }}

\newcommand{\ket}[1]{\left| #1 \right>} 
\newcommand{\bra}[1]{\left< #1 \right|} 

\newcommand{\pathToFigs}[0]{./}

\begin{document}
\begin{center}
\blfootnote{
{\hspace{-8mm}\textit{* Corresponding author}\\}
\textit{Email addresses:}  jiri.prochazka@fzu.cz, lokaj@fzu.cz,
kundrat@fzu.cz
} 
{\Large\textbf{Dependence of elastic hadron collisions 
                     on impact parameter}} \\[8mm]

Ji\v{r}\'{\i} Proch\'{a}zka$^{\text{*}}$, Milo\v{s} V.~Lokaj\'{\i}\v{c}ek and Vojt\v{e}ch Kundr\'{a}t 
\\[4mm]

\textit{
Institute of Physics of the AS CR, v.v.i., 18221 Prague 8, Czech Republic}\\

\end{center}
\vspace{6mm}

\begin{center}
\begin{minipage}{.85\textwidth}
\noindent
\textbf{Abstract}\\
{\small
Elastic proton-proton collisions represent probably the greatest ensemble of available measured data, the analysis of which may provide large amount of new physical results concerning fundamental particles. It is, however, necessary to analyze first some conclusions concerning pp collisions and their interpretations differing fundamentally from our common macroscopic experience. It has been argued, e.g., that elastic hadron collisions have been more central than inelastic ones, even if any explanation of the existence of so different process, i.e., elastic and inelastic (with hundreds of secondary particles) collisions, under the same conditions has not been given until now. The given conclusion has been based on a greater number of simplifying mathematical assumptions (done already in earlier calculations), without their influence on physical interpretation being analyzed and entitled; the corresponding influence has started to be studied in the approach based on eikonal model. The possibility of peripheral interpretation of elastic collisions will be demonstrated and corresponding results summarized. The arguments will be given why no preference may be given to the mentioned centrality against the standard peripheral behaviour. The corresponding discussion of contemporary description of elastic hadronic collision in dependence on impact parameter will be summarized and the justification of some important assumptions will be considered.} 
\end{minipage} \\
\end{center}


\noindent
\textbf{keywords:} 
{\small \KEYWORDS }

\section{\label{sec:introduction}Introduction}

The properties of fundamental particles may be established by studying experimental data obtained in their mutual collisions at different energies. The corresponding results depend strongly on their sizes and internal structures. It is evident that the characteristics in individual collisions may depend strongly also on the corresponding value of impact parameter $b$. One may expect great difference between head-on collisions at $b=0$ and at some higher value of $b$ corresponding to dimensions of colliding particles. The measured results may be significantly influenced by this parameter as the frequency of greater impact parameter values rises as $2\pi b$ practically in all hadronic collision experiments. This frequency dependence on impact parameter of initial two particle states should be, therefore, always taken into account if the experimental results are to be correspondingly interpreted. It is before all the ratio of elastic and inelastic processes that may be significantly dependent on impact parameter value. At lower collision energy values elastic hadronic scattering is as a rule the most frequent hadronic process. The results obtained at different energy values may provide important conclusions concerning space sizes and possible internal structures of the colliding particles.
  
Until now only some phenomenological models have been, however, applied to in interpreting experimental data represented by measured elastic differential cross sections. For example the description of elastic collisions of charged hadrons proposed by West and Yennie (WY) \cite{WY1968} in 1968 (having been commonly used for determination of total hadronic cross section since ISR era) has not taken impact parameter into account at all. One of the first discussion on the character of hadronic collisions in impact parameter space has been done by Miettinen in 1974 \cite{Miettinen1974} (see also \cite{Miettinen1975} from 1975). According to his calculations a rather great ratio of elastic processes should correspond to central collisions (even at impact parameter $b=0$) and the average impact parameter of elastic collisions should be smaller than that of inelastic ones; which is to be denoted as surprising in the case of matter particles. 

The central character of elastic collisions has been even more confusing due to the fact that the single inelastic diffraction seemed to be peripheral (see again \cite{Miettinen1974} or, e.g., Giovannini et al.~\cite{Giovannini1979} from 1979). Such significant difference between both the kinds of diffractive collisions may be hardly brought to agreement with other experimental data as the elastic and mentioned inelastic diffraction processes should have very similar dynamics. This kind of ''transparency'' of protons (in elastic collisions) was, therefore, denoted as a puzzling question already at that time (see, e.g., Giacomelli and Jacob \cite{giacomelli_jacob1979} from 1979).
 
It was then shown in \cite{Kundrat1981} (1981) that the central character of elastic collisions has been derived due to some assumptions in corresponding collision models that have been strongly limiting. It has been shown, too, that very different (peripheral) behaviour of elastic collisions may be obtained if one admits rather strong $t$-dependence of elastic hadronic phase that has been admitted to change only slightly.

A new more general formula for description of elastic scattering for both the Coulomb and hadronic interactions has been then derived on the basis of eikonal model in 1994 \cite{Kundrat1994_unpolarized} as an alternative to that of WY. It has allowed to take into account and further study the dependence of (elastic) hadronic collisions on impact parameter. However, even if the elastic processes prevailed significantly at higher values of impact parameter (and elastic collisions could be denoted as peripheral) the non-zero probability at $b=0$ has remained. It means that the problem of physical interpretation is to be studied and analyzed further to a greater detail.

The mentioned approach of WY for description of elastic scattering of charged hadrons which has not considered the dependence of elastic hadronic collision on impact parameter at all will be commented more in \cref{sec:WY}. The eikonal model and main formulas concerning properties of elastic processes in impact parameter space will be summarized in \cref{sec:eikonal}. The given model will be then applied to experimental data represented by measured elastic differential proton-proton cross section at 53 GeV in \cref{sec:eikonal_fitting}. Some new numerical results and characteristics derived on the basis of our earlier results \cite{Kundrat1994_unpolarized,Kundrat2002} will be presented. The given  problem of centrality or peripherality of elastic collisions will be discussed to a greater detail in \cref{sec:CEPE} where the results from \cref{sec:eikonal_fitting} obtained from the given experimental data will be compared to those published earlier by Miettinen. The analysis of experimental data in dependence on impact parameter helped significantly to identify various problems in contemporary descriptions of elastic collisions. Some of these problems will be discussed in \cref{sec:open_problems}. New probabilistic model of (elastic) collisions which is trying to address the corresponding problems and systematically take into account the dependence of elastic collisions on impact parameter will be shortly introduced in \cref{sec:model_prob} as it provides new insight on the given problems. Concluding remarks may be found in \cref{sec:conclusion}.

\section{\label{sec:WY}Approach of West and Yennie}
In the case of collisions of two protons (charged hadrons) the measured elastic differential cross section has been standardly described by complete elastic scattering amplitude $\ampl{C+N}$ (\emph{neglecting spin effects}) describing the common influence of both Coulomb and hadron interactions at all measured $t$ values as (common units $\hbar = c = 1$ used)   
\begin{equation}
\frac{\text{d} \sigma^{} (s,t)}{\text{d}t} = \frac{\pi}{sp^2}|\ampl{}|^2
\label{eq:difamp_gen}
\end{equation}
where $s$ is the square of the total center-of-mass energy and $p$ is the value of momentum of one incident proton in the center-of-mass system; $t = - 4p^2 \sin^2 \frac{\theta}{2}$ where $\theta$ is scattering angle.

According to Bethe \cite{Bethe1958} (1958) the complete amplitude has been commonly decomposed into the sum of the Coulomb scattering amplitude \ampl{C} and the hadronic amplitude \ampl{N} bound mutually with the help of relative phase $\alpha\phi(s,t)$:
\begin{equation}
  \ampl{C+N} = \ampl{C}\e^{\text{i}\alpha\phi(s,t)}+\ampl{N}
\label{eq:FCNbethe}
\end{equation}
where $\alpha=1/137.036$ is the fine structure constant. The $t$-dependence of relative phase $\alpha\phi(s,t)$ has been determined on various levels of sophistication. The dependence having been commonly accepted in the past was proposed by West and Yennie \cite{WY1968} (1968) within the framework of Feynman diagram technique (one-photon exchange) in the case of charged \emph{point-like particles} and \emph{partially in high energy limit} ($s \gg m^2$, $m$ standing for nucleon mass) as
\begin{equation}
  \phi_{\text{WY}}(s,t) = \mp \left[\ln\left (\frac{-t}{s}\right) - \int_{-4p^2}^0 \frac{\text{d}t'}{\abs{t-t'}} \left(1-\frac{F^\text{N}(s,t')}{F^\text{N}(s,t)}\right)\right].
\label{eq:phaseWY}
\end{equation}
The upper (lower) sign corresponds to the scattering of particles with the same (opposite) charges. According to \cref{eq:phaseWY} the $t$-dependence of the relative phase between the Coulomb and hadronic amplitudes may be calculated from $t$-dependent hadronic amplitude \ampl{N} entering into the integrand. In order to derive \cref{eq:phaseWY} also the Coulomb amplitude has been used (assuming a "known" form from QED, see \cite{WY1968} for details). Note that in order to derive \cref{eq:phaseWY} only a Coulomb amplitude for point-like scattering particles has been used; no form factors
have been taken into account at this stage of calculations (see \cite{WY1968} also for used phase of Coulomb amplitude \ampl{C}).

The hadronic amplitude $\ampl{N}$ may be then written using its modulus \modulus{N} and phase \phase\ as
\begin{equation} 
  \ampl{N} = \text{i}\modulus{N} \e^{-\text{i}\phase} .
  \label{eq:modphas}
\end{equation}
Formula \refp{eq:phaseWY} containing the integration over all admissible values of four-momentum transfer squared $t'$ seemed to be complicated when it was proposed. To perform analytical integration it has been assumed for the two following quantites:
\begin{itemize}
\item{the ratio of real to imaginary part of hadronic amplitude
\begin{equation}
  \rho(s,t) = \frac{\Re \ampl{N}}{\Im \ampl{N}}
  \label{eq:rho}
\end{equation}
}
\item{and diffractive slope defined as
\begin{equation}
  B(s,t) = \frac{\text{d}}{\text{d} t} \left[ \ln \dcs{\text{N}}(s,t)\right]
  = \frac{2}{\modulus{N}} \frac{\text{d}}{\text{d}t}\modulus{N}
  \label{eq:slope}
\end{equation}
}
\end{itemize}
to be $t$-independent for \emph{all kinematically allowed $t$ values}, see \cite{WY1968,Amaldi1973_231} and \cite{KL1989,KL1996}. The diffractive slope being $t$-independent means that the modulus \modulus{N} has been taken as purely exponential function of $t$. With the help of \cref{eq:modphas,eq:rho} one may obtain relation
\begin{equation}
  \tan{\phase} = \rho(s,t)
  \label{eq:tanzeta}
\end{equation}
which implies that the assumption of $t$-independent quantity $\rho(s,t)$ is equivalent to requirement of $t$-independent hadronic phase \phase.

For the relative phase between the Coulomb and elastic hadronic amplitude the following simplified expression has been then obtained for \emph{small values of $t$ only}:
\begin{equation}
   \alpha \phi(s,t) = \mp  \alpha \left [\ln{\left(\frac{-B(s)t}{2}\right)}+\gamma \right]
  \label{eq:phiWY}
\end{equation}
where $\gamma=0.577215$ is Euler constant and $B$ is $\;t$-independent diffractive slope. As introduced in \cite{KL2005,KL2007} some other high energy approximations and limitations were added, too. 

Optical theorem relating the imaginary part of elastic hadronic scattering amplitude at $t=0$ (corresponding to zero scattering angle) to total hadronic cross section
\begin{equation}
  \CS[etype=tot](s) = \frac{4 \pi}{p\sqrt{s}} \Im F^{\text{N}}(s,t=0)
  \label{eq:optical_theorem}
\end{equation}
has been then applied to and the complete elastic scattering amplitude~\refp{eq:FCNbethe} of Bethe has been written as
\begin{equation} 
\begin{split}
  \ampl[WY]{C+N} =& \pm   \frac{\alpha s}{t}f_1(t)f_2(t) \e^{\text{i}\alpha \phi(s,t)} 
                  + \frac{\CS[etype=tot](s)}{4\pi}p\sqrt{s}(\rho(s)+\text{i})\e^{B(s)t/2}.
  \label{eq:simplifiedWY}
\end{split}
\end{equation}
The two quantities $f_1(t)$ and $f_2(t)$ stand for the electric form factors. The second term on the right hand side of this equation represents the hadronic amplitude \ampl[WY]{N}. 

The Coulomb differential cross section (including form factors) has been then taken as
\begin{equation}
\frac{\text{d}\CS[etype=C](s,t)}{\text{d}t} = \frac{\pi s}{p^2} \frac{\alpha^2}{t^2} f_1^2(t)f_2^2(t),
\label{eq:dcs_c_qed}
\end{equation}
i.e., diverging at $t=0$. 
In high energy limit the Coulomb differential cross section~\refp{eq:dcs_c_qed} may be further simplified to
\begin{equation}
\frac{\text{d}\CS[etype=C](s,t)}{\text{d}t} = \frac{4\pi\alpha^2}{t^2} f_1^2(t)f_2^2(t).
\label{eq:dcs_ampl_c_high_energy_limit}
\end{equation}

The simplified formula~\refp{eq:simplifiedWY} of WY has been commonly applied to experimental data (using \cref{eq:difamp_gen}) for determination of \CS[etype=tot], $\rho$ and $B$ at various energies in very narrow interval of small values of $|t|$ (as the validity of WY approach has been limited only to these small values). The \cref{eq:simplifiedWY,eq:phiWY} derived by WY were derived in similar form also by Locher~\cite{Locher1967} even one year earlier (in 1967).

Integrated elastic hadronic cross section may be obtained from established hadronic amplitude \ampl{N} 
and \cref{eq:difamp_gen} as follows
\begin{equation} 
  \CS[etype={el}](s) = \int\limits_{t_{\text{min}}}^0 \dcs{\text{N}}(s,t) \text{d} t.
  \label{eq:sigmael} 
\end{equation}
The inelastic cross section may be then defined as

\begin{equation}
  \CS[etype=inel](s) = \CS[etype={tot}](s) - \CS[etype={el}](s).
  \label{eq:sigmainel} 
\end{equation}

It is evident that the description of elastic scattering in the approach of WY has been based on very limiting assumptions simplifying corresponding calculations without any physical motivation (e.g., the dependence of $\rho$ and $B$ on $t$ has not been considered). The dependence of (elastic) hadronic collisions on impact parameter has not been taken into account in the given approach. Any attempt has not been done to study some correlations of initial state characteristics and final state ones which should be the main goal of any experimental data analysis. 
  
In elastic collisions one should take into account mainly correlations between impact parameter values of colliding particles and angle deviations (i.e., values of $t$) of scattered particles. The first attempt in this direction has been done with the help of the eikonal model, see next section.

\section{\label{sec:eikonal}Eikonal model approach}
\subsection{Coulomb-hadronic interference formula}
One of the fundamental differences between the eikonal model approach and the approach of WY (based on Feynman diagram technique) is that the former one is trying to take into account the influence of impact parameter and establish some characteristics of hadron collisions depending on this fundamental parameter. The complex amplitude of elastic collisions of two \emph{spinless} hadrons has been expressed in the form
\begin{equation} 
  F(s,q^{2}=-t) = \frac{s}{4 \pi \text{i}} \int\limits_{\Omega_b}
  \text{d}^2 b\; \e^{\text{i}\vec{q}.\vec{b}} [\e^{2\text{i}\delta(s,b)} - 1] 
  \label{eq:Fb}
\end{equation}
where $\,\Omega_b\,$ represents two-dimensional Euclidean space of impact parameter $\,\vec{b}\,$ and $\,\delta(s,b)\,$ is so-called eikonal function. The vector $\,\vec{q}\,$ is defined as difference $\,\vec{p}-\vec{p'}\,$ of particle momenta before and after elastic scattering. In the case of elastic hadronic scattering the first (approximate) form of \cref{eq:Fb} was suggested by Glauber \cite{Glauber1959} (1959) in \emph{high energy} limit. Mathematically more rigorous derivation of elastic scattering amplitude in the impact parameter representation which respects a finite admissible region of momentum transfers at \emph{finite} collision energies was given by Adachi and Kotani \cite{PTPS.E65.316,PTPS.37.297,PTP.35.463,PTP.35.485,PTP.39.430,PTP.39.785} (1965-1968) and also by Islam \cite{Islam1968,Islam1976} (1968-1976).

The eikonal $\delta(s,b)$ may be calculated from energy-dependent spherically symmetric potential $V(s,r)$ according to \cite{Islam1967,Islam1968,Islam1976} as 
\begin{equation} 
  \delta(s,b) \sim \int\limits_{b}^{\infty}
  \frac{V(s,r)r\text{d}r}{\sqrt{r^2-b^2}}.
  \label{eq:eikonal2} 
\end{equation}
Potential $V(s,r)$ corresponds to potential between particles at momentary mutual positions during their motions and might be generally represented by a complex function\footnote{The meaning of complex potential in quantum mechanics is, however, not so straightforward as meaning of real potential in classical physics. Imaginary part is sometimes used to open (describe) inelastic scattering while real part describes elastic scattering.
}. Due to \cref{eq:Fb} the complete elastic amplitude \ampl{C+N} of two charged and spinless hadrons is fully determined by the complete eikonal $\delta^{\text{C+N}}(s,b)$. Taking into account the linearity of potential $V(s,r)$ in expression \refp{eq:eikonal2} and the \emph{additivity of Coulomb and hadronic potentials} Cahn (1982) \cite{Cahn1982} has derived an expression for complete amplitude $\ampl{C+N}$ in the eikonal model with the aim to rederive the simplified formula \refp{eq:simplifiedWY} of WY (introducing similar limitations on the elastic hadronic amplitude). The following more general formula for complete amplitude $\ampl{C+N}$ has been later derived in \cite{Kundrat1994_unpolarized} (1994)
\begin{equation}
  \ampl[\text{eik}]{C+N} = \pm \frac{\alpha s}{t} f_1(t)f_2(t) + \ampl{N}[1 \mp \text{i} \alpha G(s,t)]
  \label{eq:KLamplituda}
\end{equation}
\noindent where
\begin{equation} 
\begin{split}
  G(s,t) =& \int\limits^{0}_{t_{\text{min}}} \text{d}t' \left \{\ln \left (\frac{t'}{t} \right ) \frac{\text{d}}{\text{d}t'}[f_1(t')f_2(t')] \right.  
- \left.\frac{1}{2\pi}\left[ \frac{F^{\text{N}}(s,t')}{F^{\text{N}}(s,t)}-1\right ]I(t,t') \right \}
  \label{eq:KLampG} 
\end{split}
\end{equation}
\noindent and
\begin{equation}
  I(t,t') = \int\limits_{0}^{2\pi} \text{d} \Phi
  ''\frac{f_{1}(t'')f_{2}(t'')}{t''}.
  \label{eq:KLampI}
\end{equation}
In the last equation $t'' = t + t' + 2\sqrt{tt'}\cos{\Phi ''}$. The minimal kinematically allowed value $t_{\text{min}}$ in \refp{eq:KLampG} is given by
\begin{equation}
t_{\text{min}} = -s + 4 m^2
\label{eq:tmin}
\end{equation}
in the case of two hadrons with the same masses $m$. The upper (lower) sign in \cref{eq:KLamplituda} corresponds then to the scattering of particles with the same (opposite) charges. 

Formula \refp{eq:KLamplituda} has been derived for \emph{any $s$ and $t$ value} with the accuracy up to terms \emph{linear in $\alpha$}. It has been derived with the aim not to put any a priori unreasoned strong restriction on hadronic amplitude \ampl{N} (to avoid limitations on \ampl{N} in the WY approach). The formula~\refp{eq:KLamplituda} may be, therefore, used in two distinct ways. Firstly, it is possible to perform an analysis of measured differential cross section in the whole measured $t$-region (i.e., not only in very limited interval of small values of $t$ as it is in the case of the approach of WY) and to determine corresponding hadronic amplitude \ampl{N}. For this purpose one may choose suitable parametrizations of the corresponding modulus and phase and try to determine \ampl{N} from experimental data. Secondly, the formula \refp{eq:KLamplituda} for complete elastic scattering amplitude \ampl{C+N} may be used, too, to obtain prediction of measured differential cross section with the help of \cref{eq:difamp_gen} if the hadronic amplitude \ampl{N} has been specified within a framework of some phenomenological model description. The former usage of the formula \refp{eq:KLamplituda} will be applied to in \cref{sec:eikonal_fitting}.     

The determination of total hadronic cross section from the hadronic amplitude \ampl{N} in the eikonal model has been based on optical theorem given by \cref{eq:optical_theorem} similarly as in the approach of WY. The main advantage (and motivation) of the eikonal model over the approach of WY is that it allows to study some characteristic of collisions in dependence on impact parameter as it will be shown in the following.

\subsection{\label{sec:eikonal_profile_functions}Unitarity of $S$ matrix and $b$-dependent profile functions}
According to van Hove \cite{Hove1963,Hove1964} (1963-1964) the unitarity condition of $S$ matrix ($S^+S=1$) may be written in terms of $t$-dependent hadronic amplitude \ampl{N} as (see also \cite{Islam1968})
\begin{equation} 
  \Im \ampl{N} = \frac{p}{4\pi\sqrt{s}} \int \text{d}
  \Omega' F^{\text{N}^*}(s,t') F^{\text{N}}(s,t'') + G_{\text{inel}}(s,t)
  \label{eq:preunitarity}
\end{equation}
where $\text{d}\Omega' = \sin \vartheta'\text{d}\vartheta' \text{d}\Phi'$, $t = -4p^2 \sin^2{\frac{\vartheta}{2}}$, $t' = -4p^2 \sin^2{\frac{\vartheta'}{2}}$, $t'' = -4p^2 \sin^2{\frac{\vartheta''}{2}}$ and $\cos \vartheta'' = \cos \vartheta \cos \vartheta' +\sin \vartheta \sin \vartheta' \cos \Phi'$. Variables $\vartheta$, $\vartheta'$ and $\vartheta''$ are angles defining corresponding momentum transfers squared $t$, $t'$ and $t''$ in the center-of-mass system and $G_{\text{inel}}(s,t)$ is the so-called inelastic overlap function (introduced by van Hove) representing summation over all possible production (inelastic) states as well as the integration over all other kinematical variables.

The $t$-dependent elastic hadronic amplitude \ampl{N} is standardly expressed in $b$-space using Fourier-Bessel (FB) transformation as
\begin{equation} 
  h_{\text{el}}(s,b) =\frac{1}{4p\sqrt{s}}\int\limits_{-\infty}^0 \ampl{N} J_0(b\sqrt{-t})\text{d}t
  \label{eq:hel_standard} 
\end{equation}
where $J_0$ is Bessel function of the first kind of order zero defined as
\begin{equation}
  J_0(x)=\frac{1}{2\pi}\int\limits_0^{2\pi} \e^{\text{i} x \cos \varphi }\text{d}\varphi.
\end{equation}
Function $g_{\text{inel}}(s,b)$ has been then similarly introduced as FB transformation of $G_{\text{inel}}(s,t)$.

The unitarity condition~\refp{eq:preunitarity} is then usually expressed in impact parameter space as
\begin{equation} 
  \Im h_{\text{el}}(s,b) = |h_{\text{el}}(s,b)|^2+g_{\text{inel}}(s,b).
\label{eq:unitarity_standard}
\end{equation}
Practically all approaches of hadron-hadron scattering in the impact parameter space have been based on FB transformation~\refp{eq:hel_standard} and unitarity condition~\refp{eq:unitarity_standard}; some physical meaning has been attributed to the $b$-dependent terms in \cref{eq:unitarity_standard} using so-called profile functions $\PROF{X}(s,b)$ (X=tot, el, inel) defined as
\begin{align}
  \PROF{el}(s,b)   &\equiv 4\, |h_{\text{el}}(s,b)|^2, \label{eq:ampl_prof_el_pre}\\
  \PROF{tot}(s,b)  &\equiv 4\, \Im h_{\text{el}}(s,b), \label{eq:ampl_prof_tot_pre}\\
  \PROF{inel}(s,b) &\equiv 4\, g_{\text{inel}}(s,b).   \label{eq:ampl_prof_inel_pre}
\end{align}
The factor 4 in \cref{eq:ampl_prof_tot_pre,eq:ampl_prof_el_pre,eq:ampl_prof_inel_pre} follows from our normalization of the scattering amplitudes, see \cref{eq:difamp_gen}; the factor may be different in different conventions. This definition of profile functions has been chosen so that cross sections \CS[etype=X] may be obtained by integrating of corresponding profile functions in impact parameter space as
\begin{equation}
\CS[etype=X](s) = 2\pi\int\limits_0^{\infty} b \text{d}b\;\PROF{X}(s,b).
\label{eq:integ_cs_prof}
\end{equation}
The factor $2 \pi b$ in \cref{eq:integ_cs_prof} corresponds to the weight of initial states distinguished by impact parameter as mentioned in \cref{sec:introduction}; this factor is, therefore, not part of the profile functions. Such definition will be convenient in \cref{sec:CEPE} for easier comparison of our profile functions (obtained from analysis of experimental data in \cref{sec:eikonal_fitting}) with the earlier results of Miettinen. It should hold for the profile functions due to the unitarity condition~\refp{eq:unitarity_standard} in $b$-space 
\begin{equation}
  \PROF{tot}(s,b) = \PROF{el}(s,b) + \PROF{inel}(s,b).
  \label{eq:unitarity2}
\end{equation}

However, Adachi and Kotani \cite{PTPS.E65.316,PTPS.37.297,PTP.35.463,PTP.35.485,PTP.39.430,PTP.39.785} and Islam \cite{Islam1968} showed that no direct physical meaning can be attributed to the terms in the unitarity equation~\refp{eq:unitarity_standard} (i.e., to profile functions given by \cref{eq:ampl_prof_el_pre,eq:ampl_prof_tot_pre,eq:ampl_prof_inel_pre}) at \emph{finite} collision energies $\sqrt{s}$ due to $\Im h_{\text{el}}(s,b)$ being repesented by oscillating function. It has been shown in the quoted paper that it is necessary to distinguish integrations over physical and unphysical region of $t$ in \refp{eq:hel_standard}, i.e., $t \in\langle t_{\text{min}},0\rangle$ and $t \in(-\infty,t_{\text{min}}\rangle$ (and similarly also for $g_{\text{inel}}(s,b)$). 

The elastic hadron scattering amplitude $h_{\text{el}}(s,b)$ in the impact parameter space at \emph{finite} energies may be then defined by FB transformation of the elastic hadron scattering amplitude \ampl{N} as (see \cite{PTPS.E65.316,PTPS.37.297,PTP.35.463,PTP.35.485,PTP.39.430,PTP.39.785}, \cite{Islam1967,Islam1968,Islam1976} and \cite{Kundrat2001,Kundrat2002} for more details)
\begin{equation} 
\begin{split} 
  h_{\text{el}}(s,b) =& h_1(s,b) + h_2(s,b) \\
   =&\frac{1}{4p\sqrt{s}}\int\limits_{t_{\text{min}}}^0 \ampl{N} J_0(b\sqrt{-t})\text{d}t 
    +\frac{1}{4p\sqrt{s}}\int\limits_{-\infty}^{t_{\text{min}}}\lambda(s,t)J_0(b\sqrt{-t})\text{d}t
  \label{eq:hel} 
\end{split} 
\end{equation}
where the first term $h_1(s,b)$ represents the contribution of the FB transformation of \ampl{N} from the physical region of $t$ and the second one represents the contribution corresponding to the unphysical region of $t$. The unknown function $\lambda(s,t)$ is assumed to fulfill (obey) some conditions \cite{Islam1968}. At infinite energies $t_{\text{min}}$ is equal to $-\infty$ in \cref{eq:hel} according to \cref{eq:tmin}. The function $g_{\text{inel}}(s,b)$ has been then similarly defined as FB transformation of the inelastic overlap function $G_{\text{inel}}(s,t)$ such that integration over physical and unphysical region of $t$-values has been separated: $g_{\text{inel}}(s,b)=g_1(s,b)+g_2(s,b)$. 

To obtain non-oscillating profile functions the following definitions have been introduced (instead of \cref{eq:ampl_prof_el_pre,eq:ampl_prof_tot_pre,eq:ampl_prof_inel_pre})
\begin{align}
  \PROF{el}(s,b)   &\equiv 4\,|h_1(s,b)|^2,\label{eq:ampl_prof_el_new}\\
  \PROF{tot}(s,b)  &\equiv 4\,(\Im h_1(s,b) + c(s,b)),\label{eq:ampl_prof_tot_new}\\
  \PROF{inel}(s,b) &\equiv 4\,(g_1(s,b) + K(s,b)+c(s,b)) \label{eq:ampl_prof_inel_new}
\end{align}
and the unitarity condition at finite energies has been written in the form
\begin{align}
  \Im h_1(s,b) + c(s,b) &=  |h_1(s,b)|^2+g_1(s,b) + K(s,b) + c(s,b)
  \label{eq:unitarity1} 
\end {align}
where the real function $c(s,b)$ has been chosen so that the total and inelastic profile functions have been non-negative and main characteristics of the scattering of the scattering (like total cross section calculated from \cref{eq:optical_theorem}) have remained unchanged, see \cite{Kundrat2002} for more details. The function $K(s,b)$ is a quite negligible correction that is to be added when unitarity condition at finite energies is to be fulfilled. 

In \cite{Kundrat2001} the function $c(s,b)$ has been then parameterized and fitted to experimental data together with hadronic amplitude \ampl{N}. In the following we shall make use of a different approach as we shall choose total profile function $\PROF{tot}(s,b)$ in quite fixed (Gaussian) form as it has been usually assumed; the function $c(s,b)$ may be then determined for a given amplitude \ampl{N} on the basis of \cref{eq:hel,eq:ampl_prof_tot_new}. It is, however, convenient to introduce formulas for mean impact parameters first.

In \cite{Kundrat2002} mean-square values of impact parameter $b$ (for different collision types) have been defined as (see also some previous attempts in, e.g., \cite{PTP.35.485,Ida1962,Kundrat1981})
\begin{equation}
  \meanb[n=n,etype=X] = \frac{\int\limits_0^{\infty} b^n \; \weight(b)\PROF{X}(s,b) \text{d}b}{\int\limits_0^{\infty} \weight(b)\PROF{X}(s,b) \text{d}b}
\label{eq:integ_meanb}
\end{equation}
with $n=2$ and $\weight(b)=2\pi b$. It has been shown in \cite{Kundrat2002} that the mean-squares \meanb[n=2,etype=X] (defined by \cref{eq:integ_meanb}) of total, elastic and inelastic processes may be determined from the $t$-dependent elastic hadronic amplitude \ampl{N}. The elastic mean-square \meanb[n=2,etype=el] may be written as sum of two terms (see also \cite{Henyey1976})
\newcommand{\msint}[1][]{\ensuremath{\int\limits_{t_{\text{min}}}^0 \text{d} t {#1} \modulus{N}^2}}
\begin{equation}
  \begin{split}
    \meanb[n=2,etype=el] =& \meanb[n=2,etype=mod]+\meanb[n=2,etype=ph] \\
    =& \frac{4 \int\limits_{t_{\text{min}}}^0 \text{d} t |t| \left(\frac{\text{d}}{\text{d} t} \modulus{N} \right)^2}{\msint} 
    + \frac{4 \msint{|t|}\left( \frac{\text{d}}{\text{d} t} \phase \right)}{\msint}
  \label{eq:msel}
\end{split}
\end{equation}
where the contributions of the modulus and of the phase have been separated. The first term \meanb[n=2,etype=mod] depends only on the modulus of the hadronic amplitude \ampl{N} while the second term \meanb[n=2,etype=ph] may be influenced significantly also by the hadronic phase \phase.

Similarly, the total mean-square \meanb[n=2,etype=tot] and inelastic \meanb[n=2,etype=inel] may be evaluated with the help of \ampl{N} as
\begin{equation}
  \meanb[n=2,etype=tot] = \left. 4\left( \frac{\frac{\text{d}}{\text{d}t}\modulus{N}}{\modulus{N}} -\tan{\phase}{\frac{\text{d}}{\text{d}t} \phase}\right) \right|_{t=0},
 \label{eq:mstot}
\end{equation}
\begin{equation}
    \meanb[n=2,etype=inel] = \frac{\CS[etype=tot](s) \meanb[n=2,etype=tot] -  \CS[etype={el}](s) \meanb[n=2,etype=el]}{\CS[etype=inel](s)}.
  \label{eq:msinel}
\end{equation}
\cref{eq:msel,eq:msinel} allow comparison of \meanb[n=2,etype=inel] and \meanb[n=2,etype=el] corresponding to a given hadronic amplitude \ampl{N} and, therefore, may provide basic information whether elastic collisions are more peripheral or central than the inelastic ones, as it has been discussed in \cref{sec:introduction}.

To compare our results to those of Miettinen (see \cref{sec:introduction}) we may chose Gaussian shape of total profile function $\PROF{tot}(b)$ corresponding to the commonly assumed one as \cite{Deile:2010mv_Kundrat}
\begin{equation}
\PROF{tot}(b) = \Tilde{a}_2 \e^{-\Tilde{a}_1 b^2}
\label{eq:prof_tot_gauss_simple}
\end{equation}
where $\Tilde{a}_1$ and $\Tilde{a}_2$ are some parameters which may be expressed using \cref{eq:integ_cs_prof,eq:integ_meanb} as
\begin{align}
\Tilde{a}_1 &= \frac{1}{\meanb[n=2,etype=tot]},  \\
\Tilde{a}_2 &= \frac{\CS[etype=tot]}{\pi\meanb[n=2,etype=tot]}.
\end{align}
The total profile function $\PROF{tot}$ given by \cref{eq:prof_tot_gauss_simple} may be, therefore, determined from given values of \CS[etype=tot] and \meanb[n=2,etype=tot] using optical theorem~\refp{eq:optical_theorem} and \cref{eq:mstot}, i.e., from $t$-dependent elastic amplitude \ampl{N}. It means that using FB transformation \refp{eq:hel} of \ampl{N} and \cref{eq:unitarity2} the total, elastic and inelastic profile functions may be determined for a given \ampl{N} (e.g., from fitting experimental data as it will be done in next section).

Other shapes of the total profile function than the Gaussian one~\refp{eq:prof_tot_gauss_simple} should be admitted in principle, too; the problems concerning impact parameter picture of hadron collisions are, however, more complicated than it may be seen at first glance, see the following sections.

\section{\label{sec:eikonal_fitting}Application of the eikonal model to experimental data at 53~GeV}

The eikonal model briefly summarized in \cref{sec:eikonal} may be applied now to experimental data obtained earlier at CERN ISR at 53~GeV. The model has been applied to experimental data at this energy, e.g., in \cite{Kundrat1994_unpolarized}. The aim of our calculations is to produce now new numerical results (namely of the $b$-dependent profile functions) suitable for comparison to corresponding results of Miettinen as many contemporary descriptions of elastic scattering have been based on a quite similar approach.

\begin{figure}
\centering
\begin{subfigure}[t]{0.48\textwidth}
\includegraphics*[width=\textwidth]{\pathToFigs/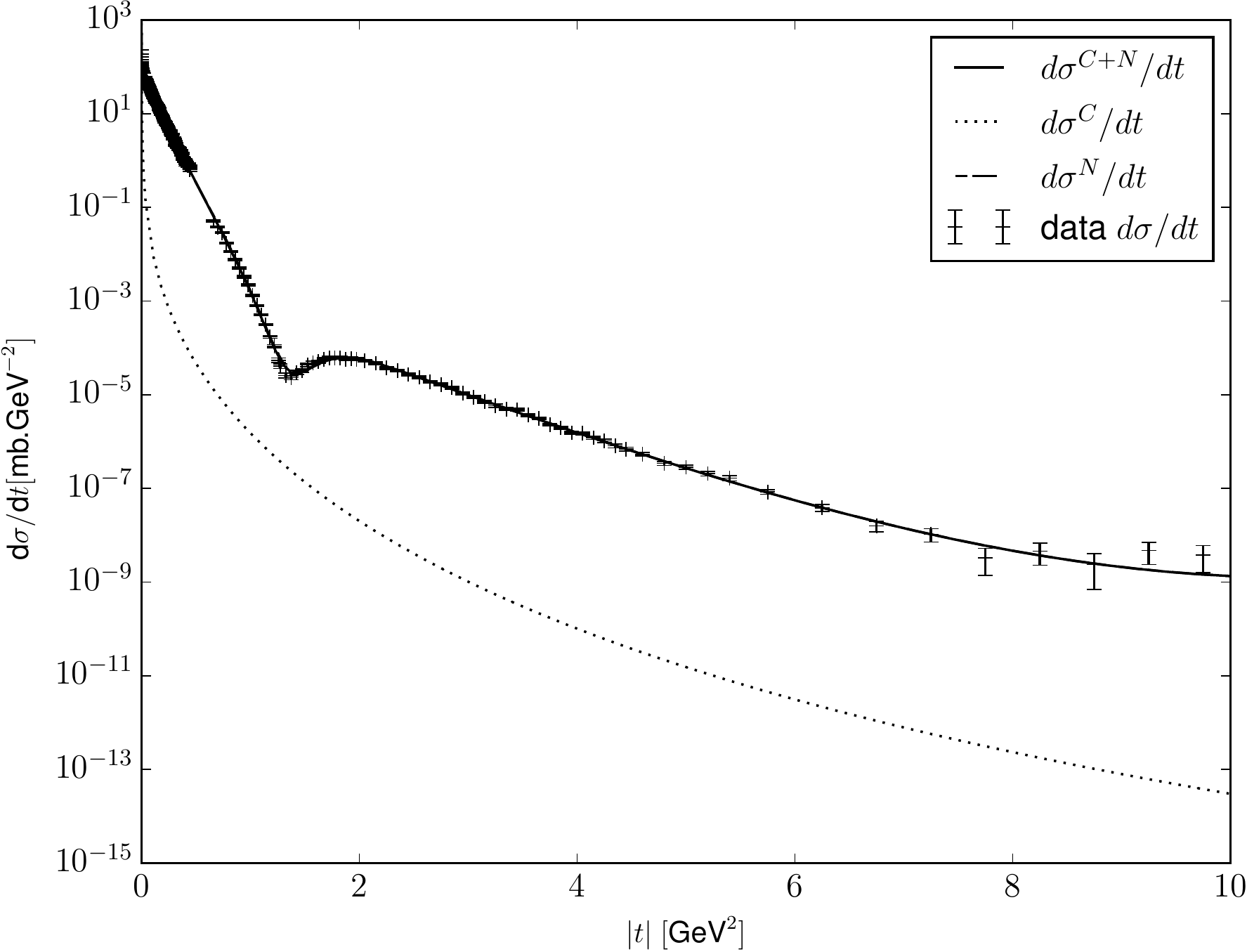}
\caption{\label{fig:dcs_53gev_peripheral_full}}
\end{subfigure}
\quad
\begin{subfigure}[t]{0.48\textwidth}
\includegraphics*[width=\textwidth]{\pathToFigs/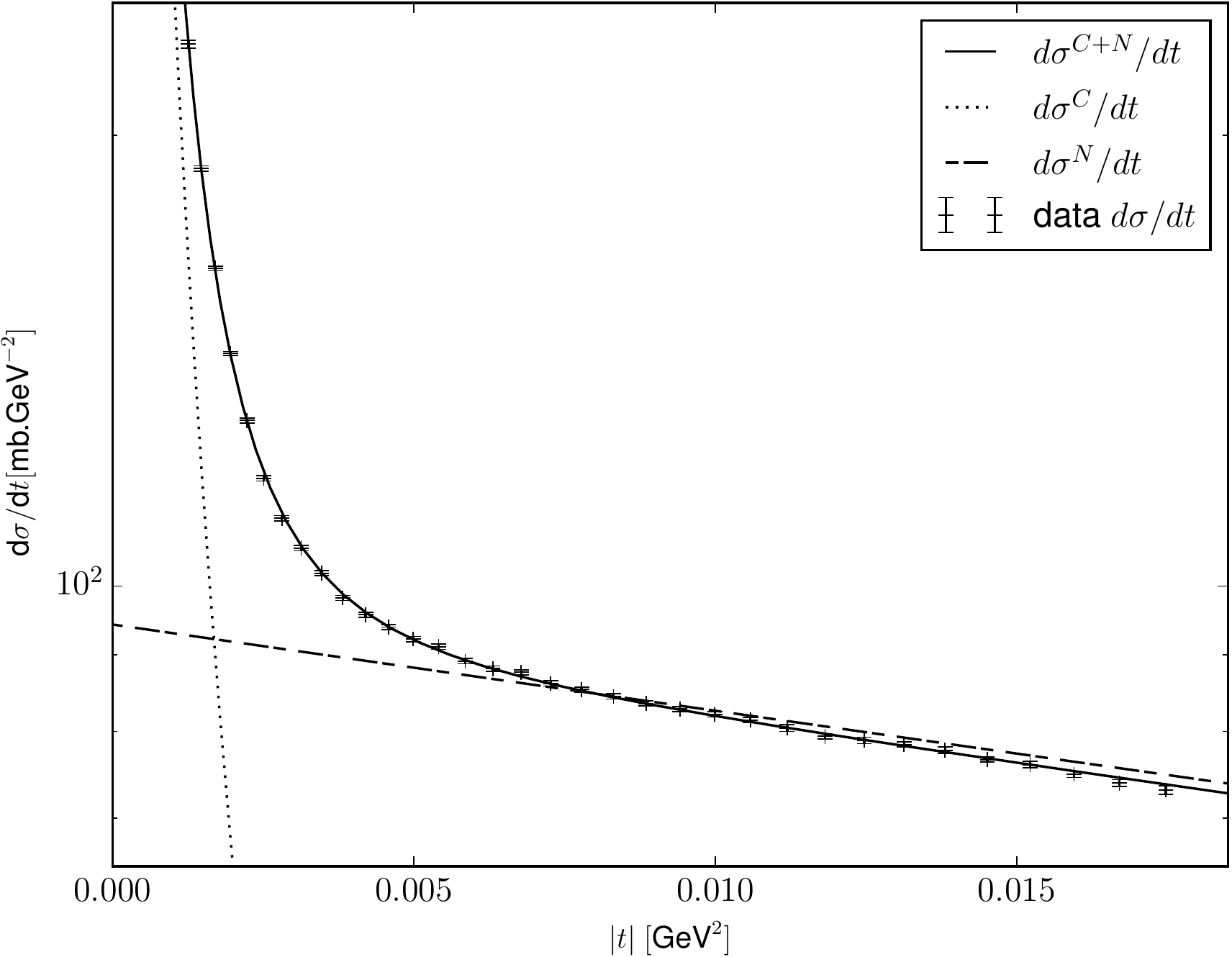}
\caption{\label{fig:dcs_53gev_peripheral_zoom}}
\end{subfigure}
\begin{minipage}[t]{.95\textwidth}
\caption{\label{fig:dcs_53gev_peripheral} Eikonal model fitted to measured pp elastic differential cross sections at energy of 53~GeV in the peripheral case (the central case having similar $t$-dependence): (a) - full fitted $t$-range, (b) - zoom to very small values of $|t|$. Individual points - experimental data, full line - Coulomb-hadronic elastic differential cross section $\text{d}\CS[etype=C+N](t)/\text{d}t$ given by eikonal model and fitted to the experimental data, dotted line - Coulomb differential cross section $\text{d}\CS[etype=C](t)/\text{d}t$, dashed line - hadronic differential cross section $\text{d}\CS[etype=N](t)/\text{d}t$.}
\end{minipage}
\end{figure}

Measured elastic pp differential cross section data at 52.8~GeV (denoted as 53~GeV) corresponding to measured range of $|t|\in \langle 0.00126,9.75 \rangle\text{ GeV}^2$ will be taken from \cite{Bystricky1980}. The data includes observed dip at $t_{\text{dip}}=-1.375\text{ GeV}^2$, see the data points in \cref{fig:dcs_53gev_peripheral_full}.

Measured pp elastic differential cross section at given energy may be analyzed using the eikonal model of Coulomb-hadronic scattering using formulas~\refp{eq:difamp_gen} and \refp{eq:KLamplituda} if the hadronic amplitude is conveniently parameterized.

To fit data in the whole measured region of $t$ the modulus of elastic hadronic amplitude may be parameterized as
\begin{equation}
  \modulus{N}= (a_1 + a_2t)\e^{b_1t+b_2t^2+b_3t^3}+(c_1 + c_2t)\e^{d_1t+d_2t^2+d_3t^3}.
  \label{eq:ampl_n_mod_param} 
\end{equation}
The corresponding phase \phase\ in \cref{eq:modphas} may be then parameterized as (see Bailly et~al.~\cite{Bailly1987})
\begin{equation} 
  \phase = \arctan{\frac{\rho_0}{1-\abs{\frac{t}{t_{\text{dip}}}}}} 
  \label{eq:ampl_n_stdphase}
\end{equation}
similarly to the $t$-dependent phase that has been used in Miettinen's calculations leading to the central bahaviour of elastic collisions in dependence on impact parameter. This parameterization reproduces the dominance of imaginary part of elastic hadronic amplitude in a rather broad region of $|t|$ and zero imaginary part at position of the dip ($t=t_{\text{dip}}$) commonly assumed in many contemporary phenomenological models (even if no reasoning of such strong and limiting assumptions has been given). Such ''standard'' form of the phase leads to central behaviour of elastic collisions as it was demonstrated for several models at the LHC energies in \cite{Kaspar2011}.

However, to study the possibility of peripheral behaviour of elastic collisions it is necessary to use more general and flexible parameterization of the phase
\begin{equation} 
  \phase = \zeta_0 + \zeta_1 \abs{\frac{t}{t_0}}^\kappa \e^{\nu t} 
\label{eq:ampl_n_zeta_gen}
\end{equation}
where $t_0 = 1 \text{ GeV}^2$. Such a parameterization may lead (according to the values of free parameters) to either central or peripheral behaviours of elastic hadronic collisions. The parametrization represents a class of functions and it has been used (with the help of corresponding penalty functions) to find the peripheral case for which $\sqrt{\meanb[n=2,etype=el]} > \sqrt{\meanb[n=2,etype=inel]}$ and elastic profile function $\PROF{el}(b)$ has maximum at $b>0$. The form factors needed in \cref{eq:KLamplituda} has been taken from Arrington's et al.~\cite{Arrington2007} (effective electromagnetic form factors).


\begin{figure}
\centering
\includegraphics*[width=0.47\textwidth]{\pathToFigs/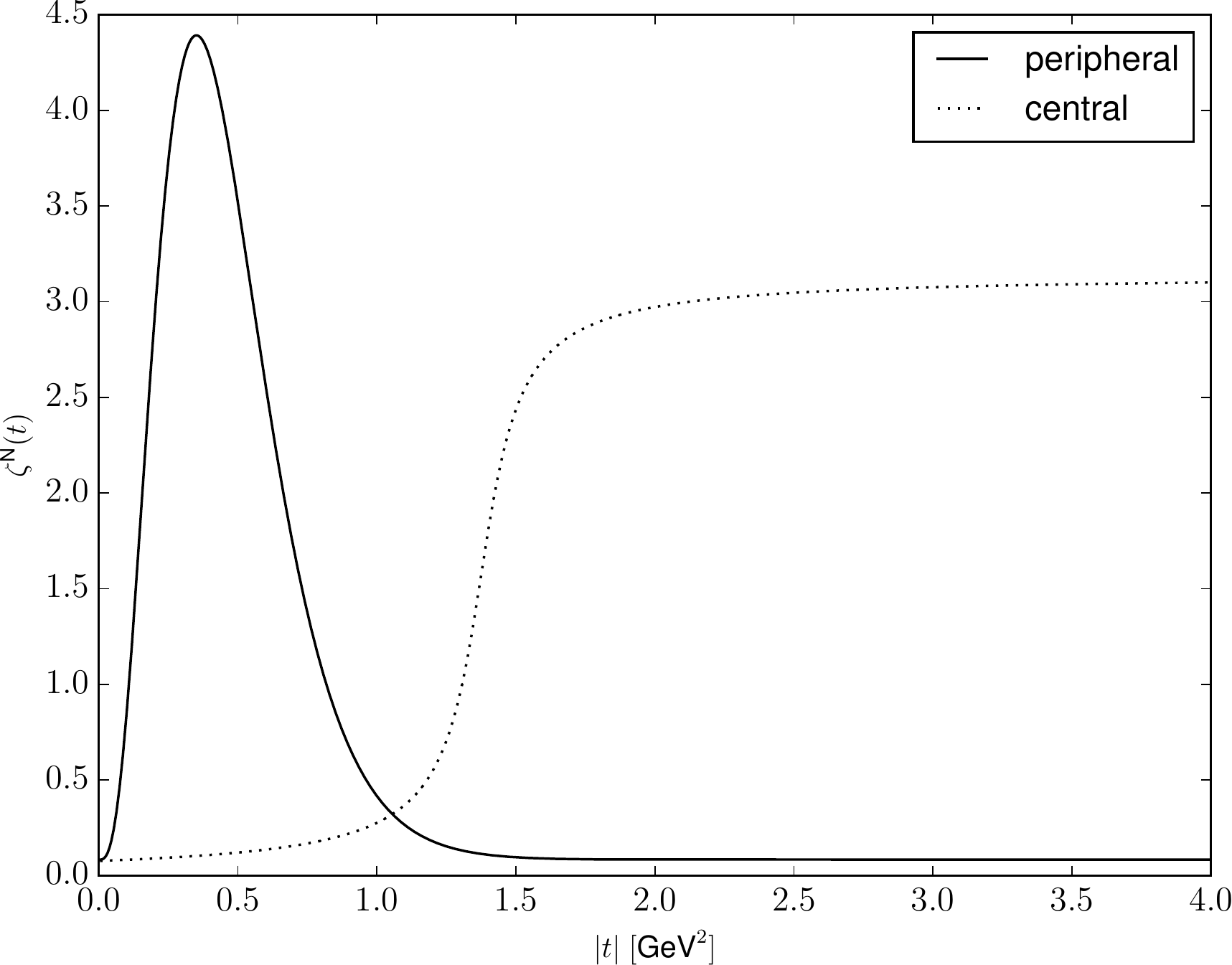} \\ 
\begin{minipage}[t]{.5\textwidth}
\caption{\label{fig:pp53gev_central_peripheral_phases}Phase of elastic hadronic amplitude \phase\ in the central and peripheral case obtained from fits of experimental data at 53 GeV.}
\end{minipage}
\end{figure}

In both the central and peripheral cases one may fit the data equally well, see the fit corresponding to the peripheral case in \cref{fig:dcs_53gev_peripheral} ($t$-dependence of hadronic differential cross section in central case at given energy having been quite similar). The values of free parameters in elastic hadronic amplitude \ampl{N} may be found in \cref{tab:model_ampl_fitted_pars}. Different $t$-dependences of hadronic phases \phase\ in peripheral and central case are plotted in \cref{fig:pp53gev_central_peripheral_phases}. Some physically interesting quantities corresponding to these two cases are shown in \cref{tab:model_ampl_quantities}. The main difference between the two fits concerns the impact parameter space as in the central case it holds $\sqrt{\meanb[n=2,etype=el]} < \sqrt{\meanb[n=2,etype=inel]}$ while the relation is reversed in the peripheral case. 
\renewcommand{\textfraction}{0.15}
\renewcommand{\floatpagefraction}{0.7}
\begin{table}
\centering
\resizebox{0.45\textwidth}{!}{
\begin{tabular}{l c c c} 
\hline \hline
fit/case         &              & central        & peripheral           \\
\hline
$\rho_0$         &              & 7.7366e-02     & -                    \\
$t_{\text{dip}}$ & [GeV$^{2}$]  & -1.375 (fixed) & -                    \\
\hline
$\zeta_0$        &              & -              & 8.3741e-2             \\
$\zeta_1$        &              & -              & 2.9709e3                \\
$\kappa$         &              & -              & 3.1930               \\
$\nu$            & [GeV$^{-2}$] & -              & 9.0967                 \\
\hline
$a_1$            &              & 1.2134e4       & 1.2178e4              \\
$a_2$            & [GeV$^{-2}$] & 1.0836e4       & 1.0870e4              \\
$b_1$            & [GeV$^{-2}$] & 5.8029         & 5.7982                \\
$b_2$            & [GeV$^{-4}$] & 3.2024         & 3.1240                \\
$b_3$            & [GeV$^{-6}$] & 1.3586         & 1.3049                \\
$c_1$            &              & 5.9472e1       & 7.0098e1              \\
$c_2$            & [GeV$^{-2}$] & -4.7858        & -2.1748             \\
$d_1$            & [GeV$^{-2}$] & 9.0217e-1      & 9.6133e-1             \\
$d_2$            & [GeV$^{-4}$] & -4.1511e-2     & -2.0792e-2            \\
$d_3$            & [GeV$^{-6}$] & -5.5806e-3     & -4.0993e-3          \\
\hline \hline
\end{tabular}}
\begin{minipage}[t]{.8\textwidth}
\caption{\label{tab:model_ampl_fitted_pars}Values of free parameters of elastic hadronic amplitude \ampl{N} in the central and peripheral case as fitted to measured proton-proton elastic differential cross section at 53~GeV.}
\end{minipage}
\centering
\resizebox{0.45\textwidth}{!}{
\begin{tabular}{l c c c } 
\hline \hline
fit/case                                       &                 & central & peripheral     \\
\hline
$\CS[etype=tot]$                               & [mb]            & 42.6    & 42.8           \\
$\CS[etype=el]$                                & [mb]            & 7.48    & 7.54           \\
$\CS[etype=inel]$                              & [mb]            & 35.2    & 35.3           \\
$\text{d}\sigma^{\text{N}}/\text{d}t(t\!=\!0)$ & [mb.GeV$^{-2}$] & 93.4    & 94.3           \\
\hline
$\rho(t\!=\!0)$                                &                 & 0.0774  & 0.0839         \\
$B(t\!=\!0)$                                   & [GeV$^{-2}$]    & 13.3    & 13.3           \\
\hline
$\sqrt{\meanb[n=2,etype=tot]}$                 & [fm]            & 1.02    & 1.02           \\
$\sqrt{\meanb[n=2,etype=el]}  $                & [fm]            & 0.678   & 1.85           \\
$\sqrt{\meanb[n=2,etype=inel]}$                & [fm]            & 1.08    & 0.728          \\
\hline
$\PROF{tot}(b\!=\!0)$                          &                 & 1.31    & 1.31           \\
$\PROF{el}(b\!=\!0)$                           &                 & 0.530   & 0.0371         \\
$\PROF{inel}(b\!=\!0)$                         &                 & 0.776   & 1.28           \\
\hline \hline
\end{tabular}}
\begin{minipage}[t]{.8\textwidth}
\caption{\label{tab:model_ampl_quantities}Values of some physically interesting hadronic quantities in both central and peripheral cases as fitted to measured proton-proton elastic differential cross section at 53~GeV.}
\end{minipage}
\end{table}

\begin{figure}
\centering
\begin{subfigure}[t]{0.42\textwidth}
\includegraphics*[width=\textwidth]{\pathToFigs/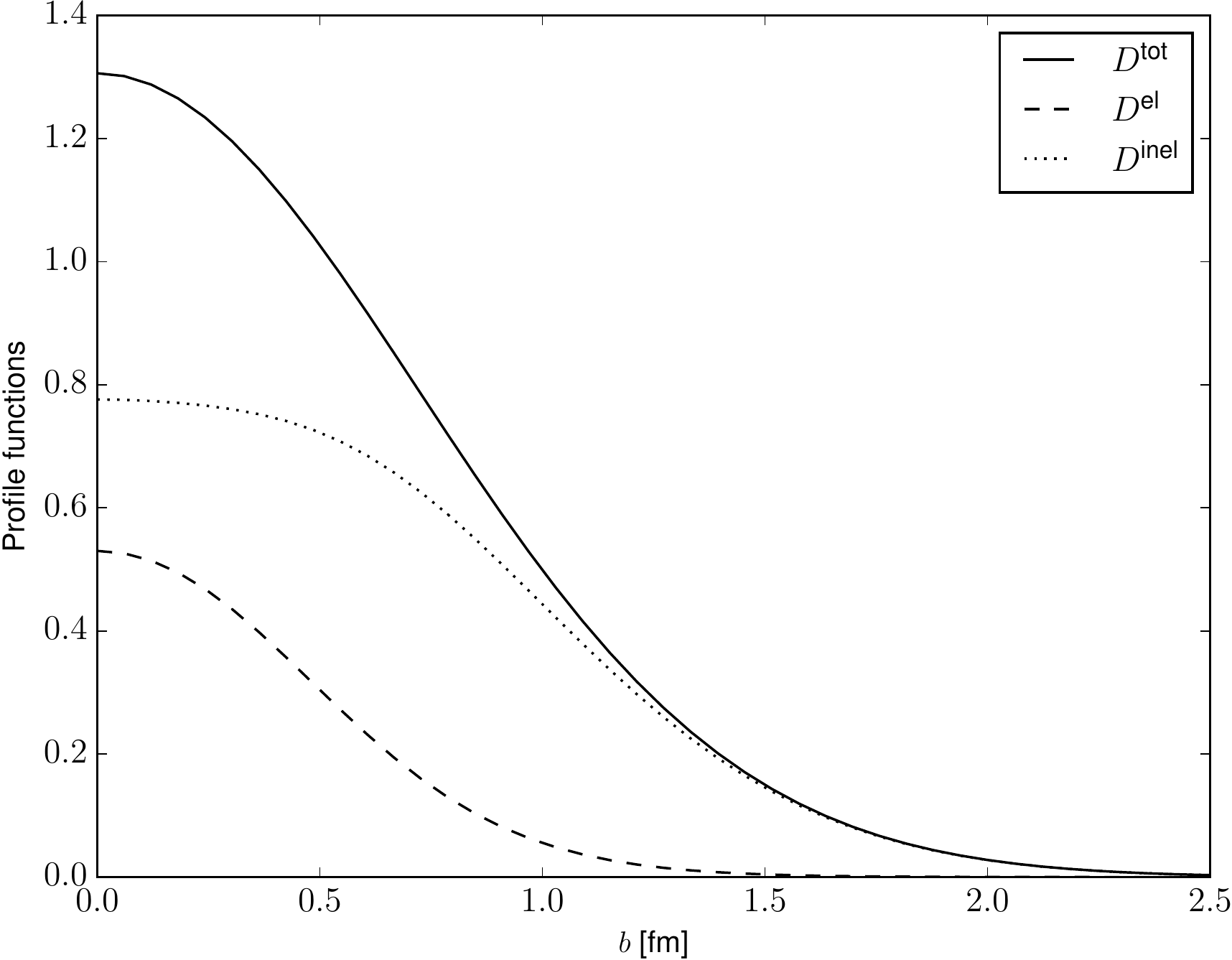}
\caption{\label{fig:profiles_53gev_central}central case}
\end{subfigure}
\quad
\begin{subfigure}[t]{0.42\textwidth}
\includegraphics*[width=\textwidth]{\pathToFigs/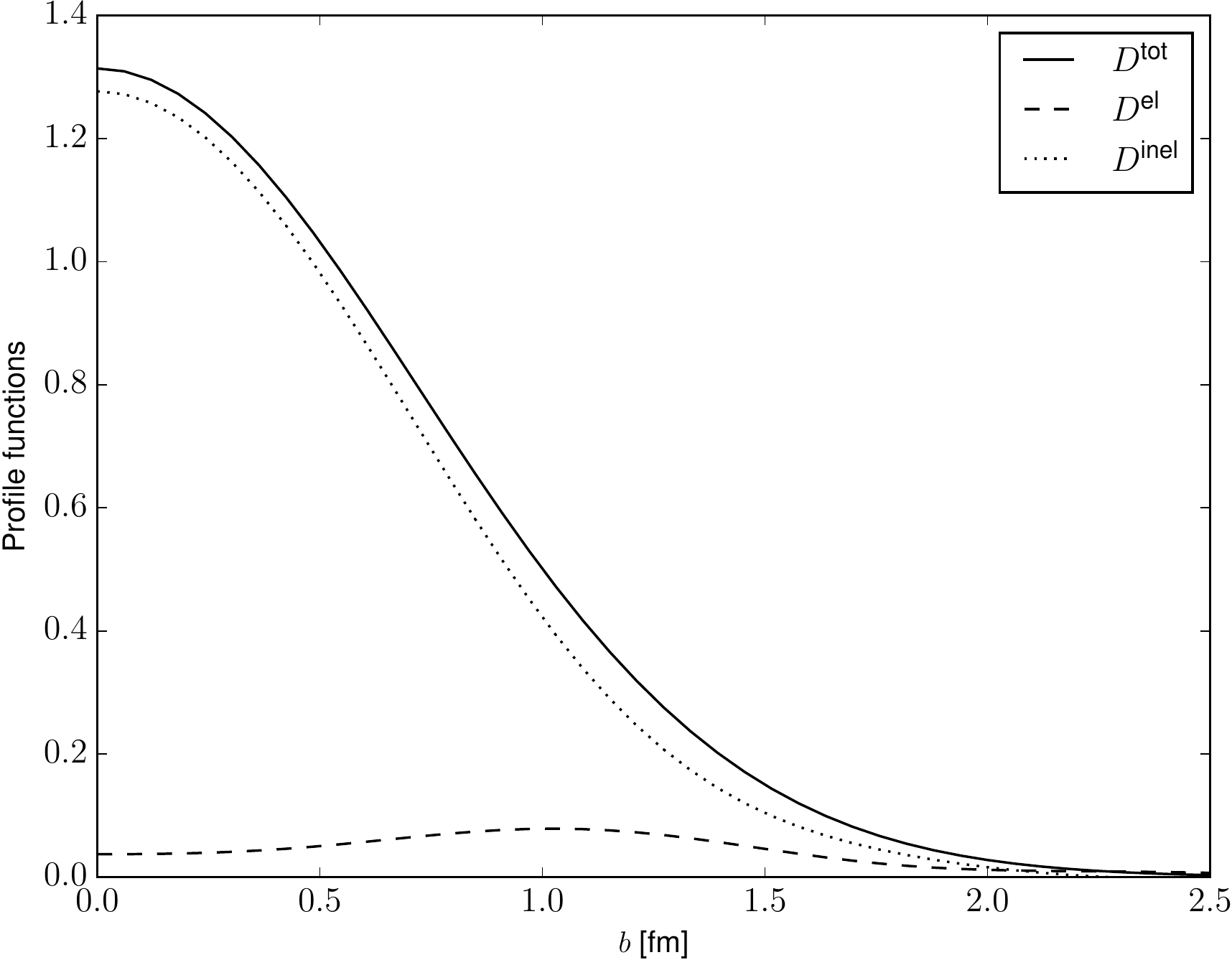}
\caption{\label{fig:profiles_53gev_peripheral}peripheral case}
\end{subfigure}
\begin{minipage}[t]{.95\textwidth}
\caption{\label{fig:profiles_53gev}Proton-proton profile functions at 53~GeV in the central and peripheral case determined on the basis of \cref{eq:unitarity2,eq:ampl_prof_el_new,eq:prof_tot_gauss_simple}. Full line corresponds to total profile function, dashed line to elastic one and dotted line to inelastic one.} 
\end{minipage}

\centering
\begin{subfigure}[t]{0.42\textwidth}
\includegraphics*[width=\textwidth]{\pathToFigs/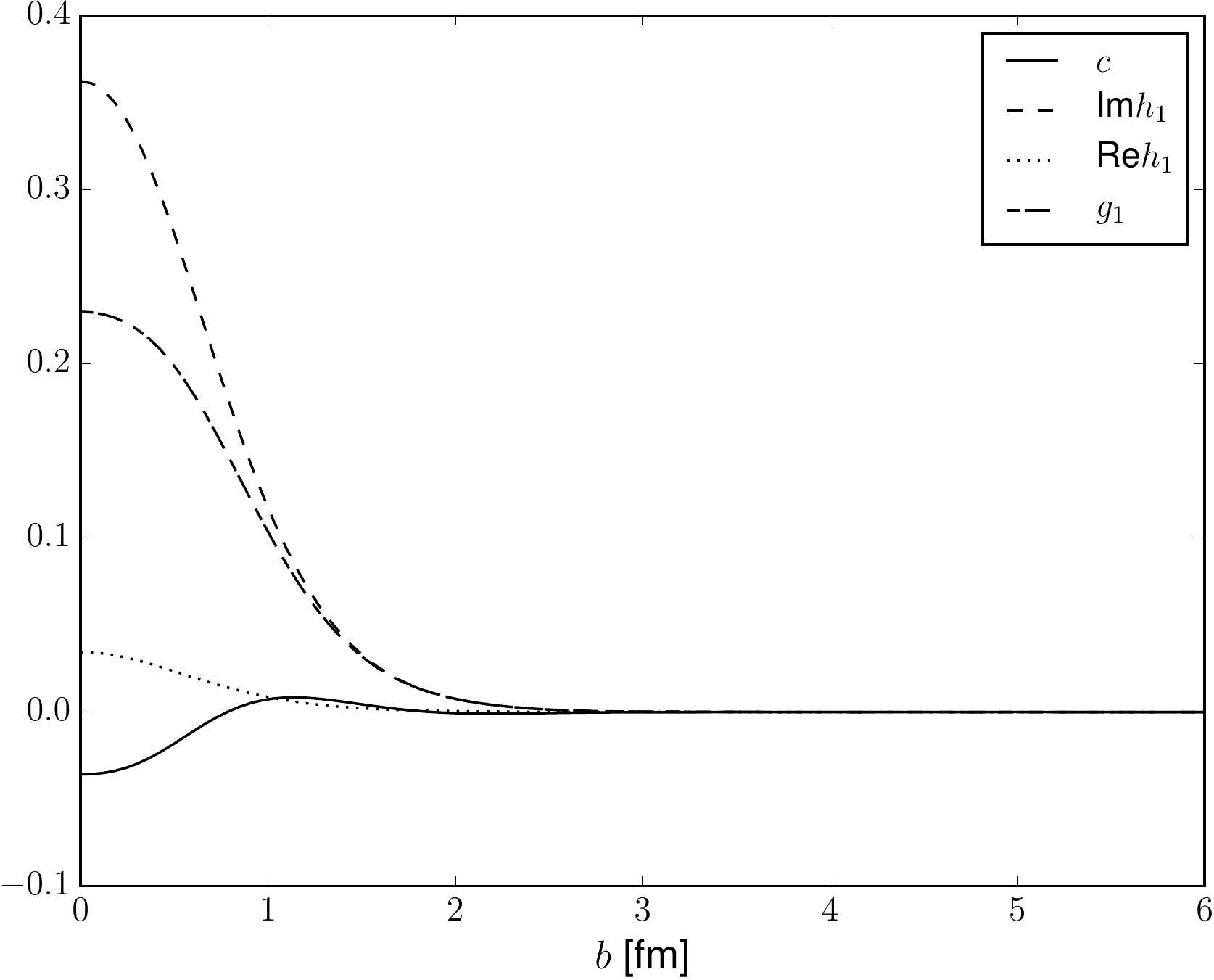}
\caption{\label{fig:noprofiles_53gev_central}central case}
\end{subfigure}
\quad
\begin{subfigure}[t]{0.42\textwidth}
\includegraphics*[width=\textwidth]{\pathToFigs/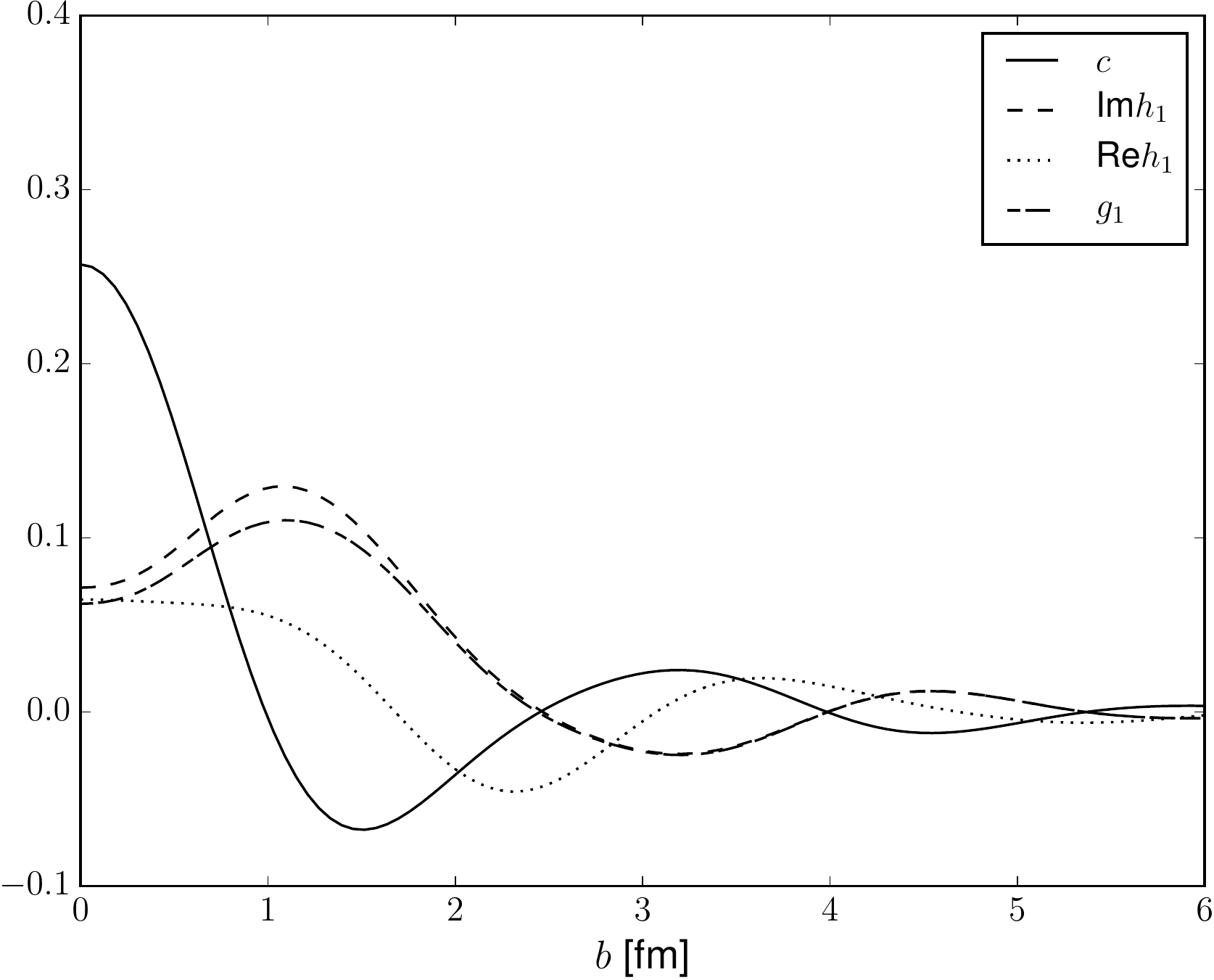}
\caption{\label{fig:noprofiles_53gev_peripheral}peripheral case}
\end{subfigure}
\begin{minipage}[t]{.95\textwidth}
\caption{\label{fig:noprofiles_53gev}Some additional functions characterizing pp collisions in dependence on impact parameter at 53~GeV in the central and peripheral case.}
\end{minipage}
\end{figure}

The difference between both the fits may be seen also in \cref{fig:profiles_53gev} where the dependence of total, elastic and inelastic profile functions on impact parameter have been plotted for both the alternatives on the basis of \cref{eq:unitarity2,eq:ampl_prof_el_new,eq:prof_tot_gauss_simple}. In the central case the value of $\PROF{el}(b\!=\!0)$ is much higher than in the peripheral case, see \cref{tab:model_ampl_quantities}; which points to big difference between elastic profile functions in both the cases. The total and inelastic profile functions plotted in \cref{fig:profiles_53gev_central,fig:profiles_53gev_peripheral} do not contain oscillations (significantly negative values) obtained in \cite{Kundrat1994_unpolarized}, see \cite{Kundrat2001,Deile:2010mv_Kundrat} for more details and similar results.
The $b$-dependent functions $\Im h_1(s,b)$, $\Re h_1(s,b)$, $g_1(s,b)$ and $c(s,b)$ in peripheral and central case are, of course, very different, see \cref{fig:noprofiles_53gev}.


From more mathematical point of view it is mainly the $t$-dependent hadronic phase which is very different in the central and peripheral case, see \cref{fig:pp53gev_central_peripheral_phases}; corresponding hadronic moduli are quite similar as they are determined practically uniquely from fitted data. The eikonal interference formula \cref{eq:KLamplituda} constrains hadronic phase \phase\ only very weekly. The parameterization given by \cref{eq:ampl_n_stdphase} leads to unique fit because the $t$-dependence is very constrained (in many phenomenological models such a priori choice of parameterization has practically predetermined the fitted results). In the peripheral case using much less constrained  parameterization~\refp{eq:ampl_n_zeta_gen} of the phase a series of different fits might be obtained (according to chosen penalty functions); only one example fit having been shown. Neither central nor peripheral fit of experimental data is, therefore, unique. It is possible to say that the ambiguity in determination of $t$-dependence of elastic hadronic phase is related to ambiguity in determination of behaviour of (elastic) collisions dependent on the impact parameter. The results following from different assumptions concerning the influence of impact parameter values should be derived and experimentally tested.

As to the form factors the addition of magnetic form factor to the electric one (used, e.g., originally in \cite{Kundrat1994_unpolarized}) does not change the character of the results as all the values of physically interesting hadronic quantities from \cref{tab:model_ampl_quantities} remain practically the same.

Quantities which may be calculated from the fitted $t$-dependent amplitude \ampl{N} and from the determined $b$-dependent profile functions may be compared to check consistency of performed calculations (for each fit separately). The total hadronic cross section given by optical theorem~\refp{eq:optical_theorem}, elastic cross section \CS[etype=el] obtained by integrating \cref{eq:difamp_gen} and inelastic cross section (given by their difference) have the same values as the values obtained from \cref{eq:integ_cs_prof}. Similarly, mean impact parameters $\sqrt{\meanb[n=2,etype=X]}$ for total, elastic and inelastic hadronic interaction calculated on the basis of \cref{eq:msel,eq:msinel,eq:mstot} are fully consistent with values obtained from \cref{eq:integ_meanb}.   

The central and peripheral fits performed in this section will be further discussed and compared to corresponding results of Miettinen in next section.

\clearpage


\FloatBarrier
\section{\label{sec:CEPE}Centrality or peripherality of elastic collisions?}
Basic experimental characteristic established in elastic collisions of protons has been represented by measured differential elastic cross section. In the case of unpolarized proton beams its $t$-dependence has exhibited  very similar structure in all cases at contemporary high energies: there has been a dip-bump or shoulder structure following the diffraction peak characterizing the behaviour at small $|t|$ (close to $t=0$) practically for all colliding hadrons \cite{Carter1986}, see the data points at 53~GeV in \cref{fig:dcs_53gev_peripheral} as an example. 
If the influence of Coulomb interaction (existing mainly in the region of very small deviations) has been separated the elastic hadronic differential cross section has been represented by the modulus of corresponding amplitude (see \cref{eq:difamp_gen}). It means that the modulus of amplitude has been strongly constrained by the given experimental data while the corresponding phase (see \cref{eq:modphas}) has remained in principle undetermined (limited only by optical theorem, imaginary part of amplitude at $t=0$ corresponding to total hadronic cross section).

It was probably the main reason why in the first analyses the phase of elastic hadronic amplitude was taken as $t$-independent in a small interval around zero; see the approach of WY \cite{WY1968} (1968) summarized in \cref{sec:WY} where the phase has been taken as $t$-independent in the whole region of kinematically allowed values of $t$. The simplified formula~\refp{eq:simplifiedWY} of WY was used as standard tool in the era of the ISR mainly for determining total (hadronic) pp cross section from measured elastic scattering at very low values of scattering angle. However, it has been shown later that several important and very limiting assumptions have been involved in the approach of WY, see \cref{sec:WY}. Most importantly, the dependence of elastic collisions on impact parameter has not been taken into account in the approach of WY at all.

The influence of impact parameter value on hadronic collisions was not considered in the first descriptions (phenomenological models) of elastic scattering, even thought the individual collision results depend surely strongly on it. As it has been mentioned in \cref{sec:introduction}, one of the first discussion concerning the influence of impact parameter and interpretation of (elastic) hadronic collisions in $b$-space has been done by Miettinen \cite{Groot1973,Miettinen1974,Miettinen1975} (1973-1975). According to his papers a rather great ratio of elastic processes should correspond to central collisions; around $6$\% of all collisions should be elastic even at impact parameter $b=0$ (i.e., head-on collisions) in the whole ISR energy range (approximately $20-60$~GeV). 

\begin{SCfigure}
\centering
  \centering
\includegraphics*[width=0.47\textwidth]{\pathToFigs/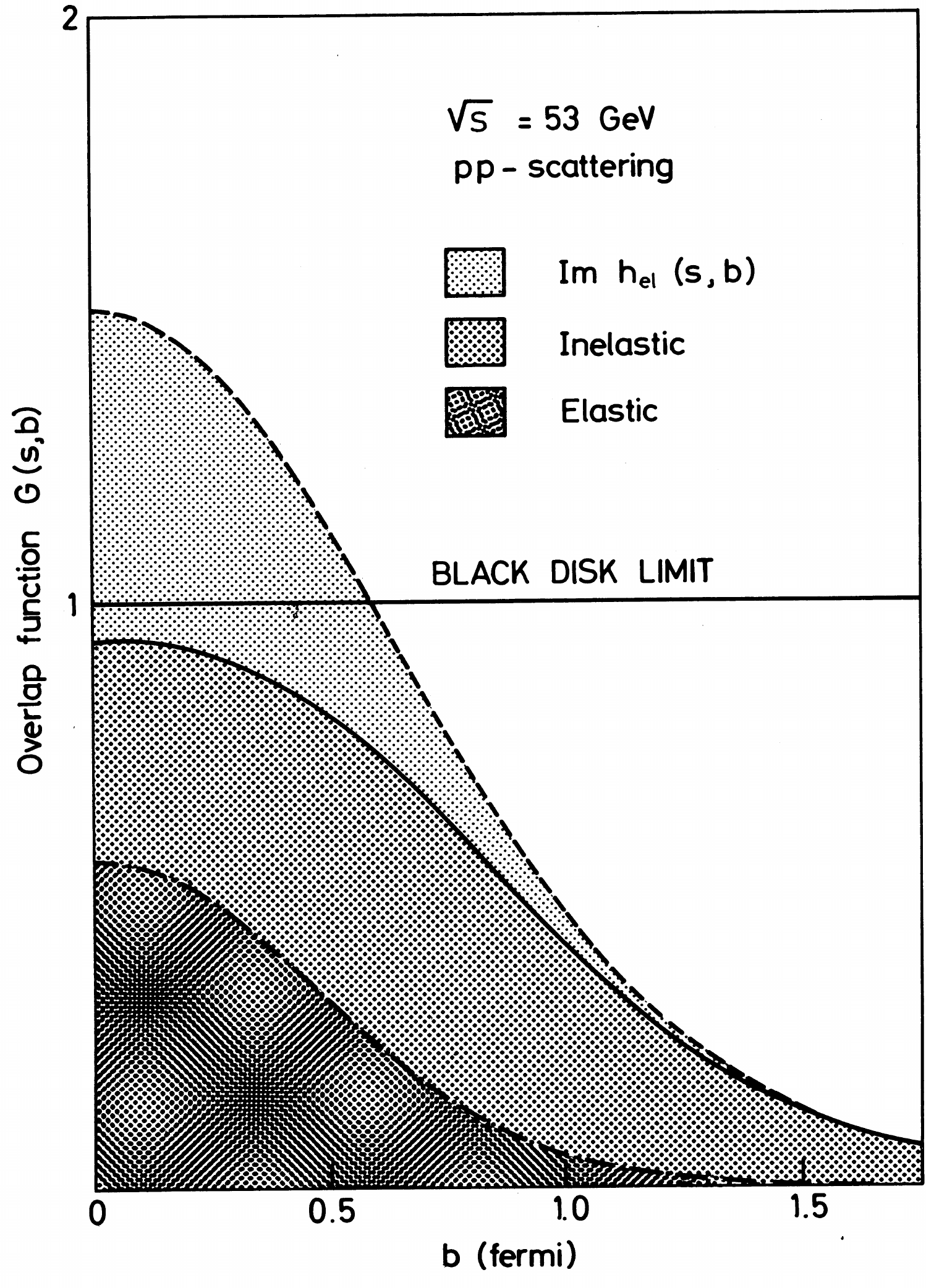}
\begin{minipage}[t]{.47\textwidth}
\caption{\label{fig:miettinen_profiles}Impact structure of pp scattering at $\sqrt{s}=53 \text{ GeV}$ derived by Miettinen \cite{Miettinen1974}. Characteristics in $b$ space are presented: imaginary part of elastic hadronic amplitude (corresponding to total profile function \cref{eq:ampl_prof_el_pre} in our notation) and $b$-dependent elastic and inelastic overlap functions (elastic and inelastic profile functions in our terminology).
 The ``black disc limit'' indicates value 1 assumed to correspond to the maximum value of inelastic overlap (profile) function allowed by unitarity (100\% absorption).}
\end{minipage}
\end{SCfigure}

The given results followed from profile functions (called $b$-dependent overlap functions by Miettinen) determined on the basis of FB transformation~\refp{eq:hel_standard} and unitarity condition~\refp{eq:unitarity_standard}. The inelastic profile function was then identified by Miettinen with inelastic probability function $\PROB[etype=inel](b)$ and subtracted from 1 (''black disc limit'' assumed to represent the maximum value of the inelastic profile function allowed by unitarity - 100\% absorption).\footnote{Definition of black disc model may be found in \cite{Svensson1967}. Colliding particles have been represented by discs of finite radius $R$ such that for impact parameters $b<R$ only absorption (inelastic) events have been allowed and for $b>R$ there has not been any hadronic interaction at all. In such a case one would expect corresponding elastic cross section being zero. However, the elastic channel has been opened in the given mathematical description when corresponding elastic scattering amplitude was not chosen to be zero; the elastic cross section in black disc model has been taken as one half of the total cross section corresponding to the disc.
}
According to \cref{fig:miettinen_profiles}, a Gaussian shape having maximum at $b=0$ and decreasing with increasing $b$ has been obtained for the profile (probability) functions of inelastic as well as elastic collisions. 

It is not, however, clear from \cite{Groot1973,Miettinen1974,Miettinen1975} why the elastic probability at $b=0$ was calculated from the inelastic one (it was put $\PROB[etype=el](b\!=\!0) = 1 - \PROB[etype=inel](b\!=\!0)$) and not directly taken from elastic profile (probability) function which was also established and which had the value of approximately 50\% at $b=0$ (which is very different value than the previous 6\%). It has not been explained, either, why total profile function being the sum of elastic and inelastic profile (probability) functions (imaginary part of elastic hadronic amplitude $\Im h_{\text{el}}(s,b)$ in Miettinen's notation, see  \cref{fig:miettinen_profiles}) has been significantly greater than 1 at values of $b < 0.6$~fm.   

The suggested central character of elastic collisions (according to \cite{Miettinen1974} p.~6: 'hadrons seem to be rather airy objects') was very confusing due to the fact that the single inelastic diffraction seemed to be peripheral (see again \cite{Miettinen1974} or, e.g., Giovannini et al.~\cite{Giovannini1979} from 1979). Such significant different physical interpretations of elastic processes and inelastic diffraction processes (mainly single diffractive) having very similar dynamics may hardly correspond to reality. This kind of ''transparency'' of protons (in elastic collisions) has been denoted as a puzzling question, too, already at that time (see, e.g., Giacomelli and Jacob \cite{giacomelli_jacob1979}, 1979); it may be hardly accepted for collisions of two matter objects when in other collision events hundreds of other fundamental particles may be produced. 
    
The result of Miettinen concerning behaviour of collisions in impact parameter space (shape of profile functions) has followed mainly from assumptions that have limited the phase $\phase$ as he used hadronic phase of the ''standard'' shape, see our comments next to \cref{eq:ampl_n_stdphase}; in his papers the used parameterizations, fitted parameters and several other details of his calculations have not been, unfortunately, explicitly mentioned. It is, therefore, not so surprising that our profile functions in the central case shown in \cref{sec:eikonal_fitting} (obtained in eikonal approach) are similar to that of Miettinen (see \cref{fig:profiles_53gev_central} and \cref{fig:miettinen_profiles}); both the approaches have been based on very similar assumptions (some differences existing but with quite minor impact on the result).

Very similar approach as that of Miettinen has been commonly used up to now by many other authors. Different parameterizations of the (''standard'') hadronic phase $\phase$ which assume (under influence of WY approach) the dominance of imaginary part of elastic hadronic amplitude at $t=0$ and in a quite wide region of higher values of $|t|$ have always been applied to. This is also, e.g., the case of recently published papers \cite{Fagundes2011,Lipari2013} where one may find (elastic) profile functions at 53~GeV very similar to those of Miettinen. As to some other contemporary descriptions of elastic scattering suggested by other authors they have been analyzed at the LHC energies in \cite{Kaspar2011}. All these models assume the ''standard'' form of the hadronic phase; all of them also lead to the centrality of elastic collisions as it has been demonstrated in \cite{Kaspar2011}.  

In \cite{Kundrat1981} (1981) it has been shown that the assumptions of $t$-independent hadronic phase ($t$-independent quantity $\rho$) and purely exponential modulus \modulus{N} ($t$-independent diffractive slope $B$) included in the simplified formula~\refp{eq:simplifiedWY} of WY have led to the given central character of elastic collisions.
It is evident that the centrality of elastic collisions has been derived on the basis of a number of assumptions (having influenced mainly the phase) that have not been sufficiently specified and reasoned; the $b$-dependence of elastic collisions having not been addressed at all.

The following question has been then solved in \cite{Kundrat1981} (in 1981), i.e., what $t$-dependence of the phase $\phase$ is to be if elastic processes might be interpreted as peripheral (or even more generally what must be changed in a given theoretical framework to obtain more acceptable behaviour of elastic collisions in dependence on impact parameter). It has been shown that peripheral interpretation of hadronic collisions may be obtained if the phase has strong $t$-dependence. 
The result concerning peripherality in \cite{Kundrat1981} has been obtained without taking into account Coulomb-hadron interference. In \cref{sec:eikonal_fitting} it has been shown that one may obtain similar result if Coulomb interaction has been included on the basis of interference formula~\refp{eq:KLamplituda}, see \cref{fig:pp53gev_central_peripheral_phases} (and also \cite{Kundrat1994_unpolarized}).

The mean impact parameters of total, elastic and inelastic processes $\sqrt{\meanb[n=2,etype=X]}$ have been defined with the help of \cref{eq:integ_meanb} where $n=2$ and $\weight(b)=2\pi b$. The factor $\weight(b)=2\pi b$ has corresponded to the assumed weight of initial two-particle states (of their impact parameter values) in corresponding experiment. Such a definition of mean impact parameters has been commonly used in the past \cite{PTP.35.485,Ida1962,Kundrat1981,Kundrat2002} as it has been assumed that the profile functions had some meaning of probabilities or distributions of impact parameter value $b$. This definition of mean impact parameter with $\weight(b)=2\pi b$ depends, however, rather strongly on the $b$-dependent frequency of produced two-particle collisions.

For discussion of centrality and peripherality of elastic collisions (reflecting different instant structures and orientations of colliding particles) it seems, however, to be more suitable to put $\weight(b)=1$ in \cref{eq:integ_meanb}. Comparison of the corresponding numerical values of mean impact parameters with $\weight(b)=2\pi b$ and $\weight(b)=1$ in the central and peripheral fit discussed in \cref{sec:eikonal_fitting} may be found in \cref{tab:model_ampl_mean_b}. One may see that the values are quite different; both the definitions have also different physical meaning.

One may also calculate in both the fits the mean impact parameters using $n=1$ instead of $n=2$ in \cref{eq:integ_meanb} (for both the weights $\weight(b)$) as it is common in mathematics for definition of mean value of random variable. The numerical values of $\meanb[n=1,etype=X]$ might be then compared to values of $\sqrt{\meanb[n=2,etype=X]}$, see \cref{tab:model_ampl_mean_b}. The values of ''mean impact parameters'' might be very different according to chosen definition. For the two fits from \cref{sec:eikonal_fitting} the mean impact parameter corresponding to elastic collisions remains lower (greater) than for the inelastic case in the central (peripheral) fit independently of the chosen definition.

\begin{table}
\centering
\begin{tabular}{l c c c c c} 
\hline \hline
fit/case                        &       & central & central  & peripheral & peripheral  \\
$\weight(b)$                    &       & 1       & $2\pi b$ & 1          & $2\pi b$    \\
\hline
$\sqrt{\meanb[n=2,etype=tot]}$  & [fm]  & 0.721   & 1.02     & 0.720      & 1.02   \\         
$\sqrt{\meanb[n=2,etype=el] } $ & [fm]  & 0.475   & 0.678    & 1.32       & 1.85   \\
$\sqrt{\meanb[n=2,etype=inel]}$ & [fm]  & 0.792   & 1.08     & 0.628      & 0.728  \\
\hline
$\meanb[n=1,etype=tot]$         & [fm]  & 0.575   & 0.903    & 0.575      & 0.902  \\
$\meanb[n=1,etype=el]  $        & [fm]  & 0.378   & 0.598    & 1.10       & 1.58   \\
$\meanb[n=1,etype=inel]$        & [fm]  & 0.647   & 0.968    & 0.521      & 0.757  \\
\hline \hline                                                                 
\end{tabular}
\begin{minipage}[t]{.8\textwidth}
\caption{\label{tab:model_ampl_mean_b} Comparison of different definitions of mean impact parameter values corresponding to total, elastic and inelastic scattering calculated on the basis of \cref{eq:integ_meanb} for $n=1,2$ and $\weight(b)=1$ or $\weight(b)=2\pi b$ in the central and peripheral fit of the pp data at 53 GeV discussed in \cref{sec:eikonal_fitting}.}
\end{minipage}
\end{table}

It has been shown in \cref{sec:eikonal_fitting} that measured elastic differential cross section at given energy may be fitted as central or peripheral according to assumptions (or parametrizations) influencing $t$-dependence of phase; the corresponding profile functions commonly interpreted as some probabilities or distribution functions of impact parameter might be very different. 

It is evident that a number of important open questions and problems exists, which should be solved before one will be able to get some really reasoned conclusions concerning the structures of corresponding particles; some of them will be discussed in next section. There is not, however, any doubt that it is the analysis based on impact parameter representation of incoming collision states that has opened quite new insight.

\section{\label{sec:open_problems} Open problems in contemporary descriptions of elastic particle collisions}

Above, we have tried to show main results concerning proton structure and corresponding characteristics derived from elastic collision data on the basis of eikonal model introduced in \cite{Kundrat1994_unpolarized}. The results of elastic collisions have been interpreted on the basis of eikonal model, of course, also in the framework of other theoretical approaches. 
In these cases the eikonal model has not been used, however, for a construction of the complete elastic scattering amplitude \ampl{C+N} (i.e., for description of common influence of both Coulomb and elastic hadronic scattering, see \cref{sec:eikonal}) but for a construction of elastic hadronic amplitude \ampl{N} only. The Coulomb effect has been usually added using the WY approach described in \cref{sec:WY}.

One of the recent attempts \cite{Petrov2002,Petrov2003,Petrov2013} to interpret elastic hadronic scattering has been done within the standard Regge pole model; it has been endeavored to determine the final hadronic effect as the sum of eikonal contributions corresponding to the exchange of different (pomerons or others) trajectories. All of the individual contributions have had a typical central Gaussian shape in the impact parameter space which has held for the resulting eikonal $\delta^{\text{N}}(s,b)$, too. Corresponding elastic profile function $\PROF{el}(s,b)$ has had, therefore, also Gaussian shape leading to central behaviour of elastic collisions in dependence on impact parameter \cite{Kaspar2011}. Hadronic amplitude \ampl{N} has been then determined on the basis of \cref{eq:Fb}. The influence of Coulomb scattering has been finally described within the standard WY approach.

The analysis of results obtained in elastic collision experiments at LHC has been done on the basis of similar Reggeon trajectories also in papers \cite{Ryskin2012} and \cite{Khoze2015}. The corresponding model has been more complex being based on three channel eikonal model trying to describe also single diffraction collision. The limitation concerning the existence of central elastic collisions has remained, as in the previous case.

Similar eikonal model approach has been used, too, in so-called QCD-inspired model of Block et al.~\cite{Block1999,Block2011,Block2012,Block2015}. Also in this case the eikonal model has been used for specifying the elastic hadronic amplitude \ampl{N} on the basis of eikonal function $\delta^{\text{N}}(s,b)$ and \cref{eq:Fb}. The interaction between hadrons has been defined in terms of the interactions of their constituents (quarks and gluons) corresponding to hadronic eikonal function. The model has been applied to elastic pp scattering at 7 TeV LHC energy \cite{Block2015}. Very similar multichannel eikonal approach has been also used in the already mentioned paper \cite{Lipari2013}.

All these models have been proposed, of course, earlier; it has been demonstrated in \cite{Kaspar2011} that they have given central elastic scattering; no attention has been devoted to this fact in original papers, even if some $b$-dependent characteristics have been calculated. In both the kinds of these eikonal approaches (Regge and QCD-inspired models) the peripheral interpretation might be undoubtedly obtained, too, if they were correspondingly generalized. 

In the corresponding experiments we need to establish the complex scattering amplitude \ampl{N} from the measured elastic differential cross section. However, from experimental data it is possible to determine practically the modulus of this amplitude only; the shape of phase has not practically any relation to experimental data, being strongly limited by other (often latent) assumptions. However, the phase dependence on parameter $b$ determines whether the collision will be central or peripheral. The given behavior has been then strongly influenced by a very limited parametrization of free (fitted) functions in the standard approach, while in the model applied to in \cref{sec:eikonal_fitting} much more broader parametrization of elastic collisions has been made use of.

However, the results derived from corresponding experimental data have been influenced also by some theoretical assumptions contradicting the experimental conditions. The attention to one mistaking assumption (admission of infinite impact parameter) has been called in \cite{totem2015} concerning the recent results at 8 TeV obtained by TOTEM experiment at the LHC at CERN. In the quoted paper the comparison of standard central behaviour of elastic collisions with the peripheral one obtained with the help of the eikonal model discussed in this paper has been presented, too. There are, of course, some other problems and open questions in contemporary theory of collision processes which will be introduced and discussed in the following.
\begin{enumerate}
\item{\textit{Coulomb interaction and experimental conditions}\\
In experimental analysis it is always necessary to ''subtract'' the effect of Coulomb interaction of charged particles which is currently done under assumptions being in partial disagreement with corresponding experiments. 
\begin{enumerate}
\item{\textit{(Non)divergence at $t=0$}\\
The Coulomb differential cross section has always been assumed to be given by \cref{eq:dcs_c_qed}, i.e., diverging at $t=0$. Contemporary Coulomb-hadronic interference formulas include integration over the corresponding diverging $t$-dependence of Coulomb amplitude (which has been also the case of both \cref{eq:simplifiedWY,eq:KLamplituda}). However, such a dependence does not correspond to conditions in any relevant experiment as the singular point $t=0$ may exist only for events corresponding to infinite impact parameter value while in real experiments corresponding values are to be less than micrometers.}
\item{\textit{Multiple collisions}\\
In real experiments very small deviations (scattering angles) should be described rather as the sum of multiple scattering at higher (yet possible) impact parameters and of standard Coulomb scattering of two colliding particles. It means that also the $t$-dependence of elastic hadronic cross section in the region of the smallest deviations might be strongly influenced by subtraction of the given Coulomb interaction effect. The effect of multiple scattering was discussed already by Rutherford, Geiger and Marsden when they started to interpret first fixed-target experiments in 1911, see \cite{Rutherford1911}. It has been taken into account in some other analyses of fixed-target experiments in order to determine corresponding characteristic of only single scattering. It should be considered also in colliding-beam experiments (evaluate how much the effect may influence the (elastic) collision result at given energy).
}
\item{\textit{Electromagnetic form factors}\\
The dependence of Coulomb interactions on impact parameter has not been satisfactorily taken into account probably not only in elastic pp scattering but also in elastic ep scattering for determination of electromagnetic proton form factors (e.g., in \cite{Arrington2007} the influence of impact parameter on particle collisions not having been mentioned at all, either).}
\end{enumerate}
}
\item{\textit{Different mechanism of Coulomb and strong forces}\\
One should ask, too, whether the standardly used Coulomb-hadronic interference formulas have described the given collision data in full agreement with reality when the corresponding forces have had very different characteristics. The Coulomb force acts at any distance being efficient practically during the whole evolution time while the hadronic force should be interpreted rather as a contact force being efficient for a very short time interval at very small impact parameter values only. One should, therefore, ask how much the contact hadronic interaction may be influenced by earlier Coulomb interaction acting at greater distances.}
\item{\textit{Weak interaction}\\
The existence of some weak forces (their range being probably limited similarly as for strong ones) has been standardly considered in the description of fundamental particles but until now the elastic hadron collisions have been interpreted as superpositions of strong and Coulomb interactions only. Considering corresponding problems in a much broader context (see, e.g., probable dimension of hydrogen atom and the distances of hadronic objects in solid substances \cite{Lokajicek2013_intech}) one may come to the conclusion that the proton might interact also weakly at greater distances, the strongly interacting matter being surrounded by other matter interacting only weakly. One should, therefore, ask whether the hadronic collisions are not influenced more by such a weak interaction than by Coulomb force.
}
\item{\textit{Properties of $S$ matrix and structure of Hilbert space}\\
It is also the usual definition of $S$ operator that should be newly analyzed and tested. If the collision processes are to be represented in a Hilbert space correctly the initial and final states should be represented by vectors in two mutually orthogonal subspaces (see, e.g., \cite{Lax1976_1,Lax1976_2} and also \cite{Alda1974,Lokajicek2012,Lokajicek2013_intech,Lokajicek2014_bell}). The given Hilbert structure has been, however, excluded by Bohr in 1927 \cite{Bohr1928} who asked for the Hilbert space representing the evolution of any physical system to be spanned always on one basis of Hamiltonian eigenvectors (instantaneous states of incoming and outgoing particle pairs having always been represented by the same vector). The $S$ matrix in Bohr's Hilbert space may, therefore, hardly represent the transition probabilities from initial state to final one
\begin{equation}
P_{i\rightarrow f} = \left|\bra{f}S\ket{i}\right|^2
\end{equation}
as it is usually assumed. There is also problem with the interpretation of $S$ matrix unitarity. It will be probably necessary to define a transition operator between incoming and outgoing states (i.e., between two mutually orthogonal Hilbert subspaces) caused by a ''contact'' force (between two mutually orthogonal Hilbert subspaces); however, it will be hardly possible to ask for it to be unitary (see \cite{Prochazka2015_OT}). In commonly used approaches initial two-particle states have not been distinguished in Hilbert space according to impact parameter. Corresponding distinction of initial characteristics would allow studying transition (collision) probabilities to much more details.
}

\item{\textit{Optical theorem}\\
All contemporary models of elastic hadron collisions have been based on optical theorem validity which correlates the total cross section to the value of imaginary part of collision amplitude at one point ($t=0$), see \cref{eq:optical_theorem}. The theorem has been taken from optics and applied to strong interaction collisions while any proof of its validity in strong interactions (particle physics) has not been given until now, see \cite{Prochazka2015_OT} for detailed discussion. It has been shown in \cite{Prochazka2015_OT} that the attempts to prove optical theorem in particle physics assume unitarity of $S$ matrix (given by \cref{eq:preunitarity}) and initial state(s) not properly distinguished according to impact parameter as mentioned in the preceding paragraph. Another problem is related to the fact that the so-called ''non-interacting'' final states have been arbitrarily interchanged with elastic states at $t=0$. Both the cases correspond to zero scattering angle but the corresponding states have completely different frequencies in experiments, which has not been respected and taken into account. Detailed discussion of these problems and some others may be found in \cite{Prochazka2015_OT} where one may find also historical context (see also \cite{Newton1976}). The impact of the constraining optical theorem in contemporary models is not quite clear as it is accompanied by other assumptions. It has, however, strongly influenced contemporary extrapolations of elastic hadronic amplitude to $t=0$ since its very first application to experimental data. One should, therefore, look for description without this limitation and make corresponding comparisons.
}
\item{\textit{Determination of $b$-dependent probability functions of hadron collisions}\\
If the colliding objects are not spherically symmetrical and randomly oriented in space during collision then one may expect that the particles collide elastically at a given value of impact parameter with some probability $\PROB[etype=el](b)$. Miettinen tried to identify (without any justification) profile functions with the corresponding probabilities but with unsatisfactory result. Even if profile functions (or other $b$-dependent quantities) have been calculated by many authors up to now it has been quite rare to find in corresponding papers at least some comments related to the actual meaning of these functions. Their relation to clearly defined $b$-dependent probabilities of corresponding collisions has not been given, which has strongly limited the possibilities of interpreting measured collision data, see \cref{sec:CEPE}. 
}

\item{\textit{Distribution of elastic scattering angles for a given impact parameter}\\
In description of elastic collisions one should take into account correlations between impact parameter value of colliding particles and angle deviations (values of $t$) of scattered particles, i.e., correlations between initial and final elastic states. Two particles may interact elastically at impact parameter $b$ with probability $\PROB[etype=el](b)$ and be scattered with a value of $t$ in (rather broad) $t$-range. The corresponding $t$-distribution at each value of $b$ may be denoted as $\dbt$. This $t$-distribution is different from that corresponding to $t$-dependent differential cross section~\refp{eq:difamp_gen} (elastic hadronic amplitude \ampl{N}) as the later one does not correspond to only one value of $b$. Standardly used FB transformation \refp{eq:hel_standard} and \refp{eq:hel} (connecting $t$-dependent elastic hadronic amplitude with the $b$-dependent one of unclear physical meaning) may, therefore, hardly provide basis for determination of the functions $\PROB[etype=el](b)$ and $\dbt$.
}
\end{enumerate}

It is evident, due to the mentioned problems, that the usually applied theory of elastic collisions has remained practically in its initial stage. Many assumptions have been made use of to obtain agreement with some experimental data without testing their conformity with other experimental observations. Any concrete assumptions concerning the structures of colliding objects have not been practically proposed and tested.  


Any progress in the given area may consist evidently in abandoning mere mathematical assumptions (often contradicting experimental conditions) and in analyzing systematically the dependence of corresponding collisions on impact parameter. Assumptions without clear physical meaning (arbitrarily chosen  like, e.g., a priori assumed $t$-dependence of elastic hadronic amplitude) should be substituted by assumptions with direct relation to corresponding possible particle structures; purely mathematical-phenomenological descriptions should be substituted by ontological models of physical reality, see \cite{Lokajicek2007}. The outcome of collision models fitted to experimental data should have clear relation to particle structure while, e.g., the quantity $\rho(t=0)$ commonly calculated up to now under influence of the approach of WY (see \cref{sec:WY}) does not have such property. 
Any model of elastic collisions (even if based possibly on very different assumptions) should analyze collisions in dependence on impact parameter and be able to determine $b$-dependent probability function $\PROB[etype=el](b)$ and the $t$-distribution $\dbt$ that might be further discussed as they strongly reflect the structure of colliding particles. It is necessary to evaluate newly all assumptions concerning description of elastic scattering, which includes the commonly applied optical theorem, too.

In the next section a very preliminary version of a model corresponding to the mentioned requirements will be presented. It is based on simple assumptions concerning some $b$-dependent collision characteristics that might be expected on the basis of ontological approach that has represented the backgrounds of classical physics.

\section{\label{sec:model_prob} Probabilistic model of hadron collisions} 

One of the main problems of contemporary descriptions of elastic scattering has been related to the fact that the influence of impact parameter has not been systematically taken into account. Some $b$-dependent functions have been determined but their physical meaning have been unclear. 

To overcome at least some problems summarized in the preceding section new hadron collision model has been recently proposed by us in \cite{Lokajicek2013_intech}; it may be applied directly to elastic scattering of hadrons. The effect of Coulomb and strong forces (due to their significant difference) may be practically separated and it is not necessary to define it as the sum of corresponding amplitudes; one may sum directly corresponding probabilities. Starting from the ontological interpretation of colliding matter objects one should expect that not only the probability of collision processes at given energy but also that of elastic and inelastic collisions will depend  on the values of mutual impact parameter. One may then assume that the colliding objects are non-spherical and differently oriented in space. For the probability of elastic hadronic collision at a given impact parameter $b$ it may be then written 
\begin{equation}
   \PROB[etype=el](b) = \PROB[etype=tot](b)\, \PROB[etype=rat](b) 
\end{equation}
where $\PROB[etype=tot](b)$ is the probability of all possible hadronic (elastic or inelastic) collision processes and $\PROB[etype=rat](b)$ is the ratio of elastic probability to total one. Non-zero probabilities of short-ranged (contact) strong interactions existing only for $b\in \langle 0, \bmax \rangle$.

On the basis of ontological realistic approach one may further expect that elastic collisions should be mainly peripheral. The functions $\PROB[etype=tot](b)$ and $\PROB[etype=rat](b)$ may be then assumed to be monotonous functions of $b$: the first one diminishing with rising $b$ and the other increasing in the same interval of $b$. The probability of inelastic collision at given impact parameter may be defined as (conservation of probability)
\begin{equation}
\PROB[etype=inel](b) =  \PROB[etype=tot](b) - \PROB[etype=el](b).
\label{eq:prob_inel} 
\end{equation}

It is then necessary to admit (for non-spherical particles) that the initial state characterized by a given $b$ value may pass in the case of elastic processes to a continuous set of $t$ values. The corresponding distribution of $t$ values will have maximum at certain value of $t(b)$; in the first orientation approximation this value may be then taken as the only possibility (the $t$-distribution $\dbt$ mentioned in preceding section is given by corresponding Dirac delta function in this special case). One may then assume that elastic scattering angle is decreasing with increasing value of $b$; 
the impact parameter should be then represented by corresponding opposite function   $b(t)$. In such a case it is possible to write for elastic hadronic differential cross section
\begin{equation}
    \frac{\text{d}\CS[etype=N]}{\text{d}t} = 2\pi\,b(t)\, \PROB[etype=el](b(t)) \frac{\text{d}b}{\text{d}t}
\end{equation}
where the factor $2 \pi b$ represents frequency (weight $\weight(b)$) of initial collision states distinguished by impact parameter as colliding particles may be assumed to be distributed uniformly in cross plane. This assumption represents good approximation in the case of short-ranged hadronic interactions in collision experiments where non-zero collision probabilities correspond to impact parameters less than several femtometers.
The contemporary contribution of Coulomb interaction will be practically negligible in comparison to strong interaction contribution; it may be, however, significant for $b > \bmax$. The contribution of single interactions may be easily calculable, while the contribution of multiple collisions is to be represented by a phenomenological expression.

The integrated total, elastic and inelastic hadronic cross sections may be calculated from corresponding $\PROB[etype=X](b)$ probability functions as ($X$=tot, el, inel)
\begin{equation}
        \CS[etype=X] \;=\; \int_0^{\bmax} \text{d}b \,2\pi\,b\,\PROB[etype=X](b); 
\label{eq:integ_cs_prob}
\end{equation}
where parameter $\bmax$ corresponds to the maximum range of strong interaction.\footnote{Some assumptions of this new collision model
concerning the $b$-dependent probabilities has been formulated on the basis of ontological approach. It is interesting to compare \cref{eq:prob_inel,eq:integ_cs_prob} 
to \cref{eq:unitarity2,eq:integ_cs_prof} for $b$-dependent profile functions as they have the same form. The $b$-dependent profile functions calculated on the basis of unitarity condition and FB transformation given by \cref{eq:unitarity_standard,eq:hel_standard} have many properties as the $b$-dependent probability functions but the profile functions still cannot be easily identified with the probabilities as it has been discussed in previous sections.}

This collision model has been already applied by us to experimental data represented by measured elastic proton-proton differential cross section at energy of 53~GeV \cite{Lokajicek2013_intech} (the data used in \cref{sec:eikonal_fitting}). Even if the form of the model has been only preliminary (it has not addressed all the problems and suggestions mentioned in \cref{sec:open_problems}) it has been possible to demonstrate explicitly that new interpretation possibilities of structure and interaction of fundamental particles have been opened on its basis. Our orientation results have indicated for example that a hadron might exist in some internal states differing very slightly in their dimensions; 
for more details see \cite{Lokajicek2013_intech}.

\section{\label{sec:conclusion}Conclusion}

It has been shown that the often proclaimed centrality of elastic collisions between fundamental particles has followed from a series of assumptions that may hardly correspond to situation in matter reality. The approach of WY has influenced significantly many recent descriptions of elastic scattering in interpreting the $b$-dependence of elastic collisions even if the corresponding dependence on impact parameter has not been tested and taken into account originally in the corresponding approaches.

The peripheral characteristics of elastic collisions having been reached with the help of the eikonal model (see \cref{sec:eikonal_fitting}) do not seem, however, to represent a final solution, either. The eikonal model has allowed to study some characteristics in dependence on impact parameter but the relation of some $b$-dependent functions (e.g., profile functions) to corresponding $b$-dependent probabilities $\PROB[etype=el](b)$ and $t$-distribution for a given impact parameter $\dbt$ has failed to be established.  

It is possible to say that some conclusions concerning fundamental particles have been influenced by general attitude of physical community to the microscopic region of physical reality, when the ontological approach has been refused in the end of the first half of the past century and two quite different theories have started to be applied to physical reality (classical theory in macroscopic region and Copenhagen quantum alternative in microscopic one); the corresponding standpoint being supported now on the basis of Bell inequality violation only (for necessary details see \cite{Lokajicek2012}). However, until now no attempt pointing to the boundary between corresponding different regions of matter world has been presented. In addition to, a series of papers have been already published arguing on one hand that Bell's inequality has contradicted the conditions derived by Boole for probability values in corresponding probability experiments (see, e.g., \cite{Rosinger2004}), and on the other hand that Bell's inequality has been in contradiction to the experiment where it has been violated (see, e.g., \cite{Lokajicek2012,Lokajicek2013_intech,Lokajicek2014_bell}). There is not practically any argument against the application of ontological approach of Aristotle (applied quite successfully in classical physics) to any known matter objects \cite{Lokajicek2007}.  

As to possible progress in the research of fundamental particles main source of new pieces of knowledge may be expected from the study of elastic collisions. More general models addressing the problems considered in \cref{sec:open_problems} and allowing to test some new assumptions concerning the influence of impact parameter values on elastic collisions should be looked for. New view and new possibilities will be opened by comparing the results obtained with the help of corresponding models to the contemporary results. Some first steps in this direction may be found in \cite{Lokajicek2013_intech} as it has been mentioned in \cref{sec:model_prob}.

\end{document}